\documentclass[a4paper,fleqn]{cas-sc}

\ExplSyntaxOn
\cs_gset:Npn \__first_footerline:
  { \group_begin: \small \sffamily \group_end: } 
\ExplSyntaxOff 

\ExplSyntaxOn
\cs_gset:Npn \__first_headerline:
  {
  }
\ExplSyntaxOff

\usepackage{amssymb}
\usepackage{enumerate}
\usepackage{mathtools}
\usepackage{graphicx}
\usepackage{caption}
\usepackage{subcaption}
\usepackage{multirow}
\usepackage{adjustbox,lipsum}
\usepackage{booktabs}
\usepackage{tabularx}
\usepackage{tablefootnote}
\usepackage{cleveref}
\usepackage{capt-of}
\usepackage[sort, numbers]{natbib}
\usepackage{enumitem}
\usepackage{placeins}
\usepackage{fancyhdr}
\usepackage{placeins}

\begin{document}
\let\WriteBookmarks\relax
\def\floatpagepagefraction{1}
\def\textpagefraction{.001}

\shorttitle{Opportunity-Centric Accessibility for Public Charging Infrastructure}

\shortauthors{Gazmeh et~al.}

\title[mode=title]{Understanding the Opportunity-Centric Accessibility for Public Charging Infrastructure}                      

\author[1]{Hossein Gazmeh}
\credit{Conceptualization, Methodology, Formal analysis, Writing – Original, Review \& Editing, Visualization}
\author[2]{Yuntao Guo, Ph.D.}
\credit{Writing – Review \& Editing, Data Curation}
\author[1]{Xinwu Qian, Ph.D.\corref{cor1}}
\cortext[cor1]{E-mail: xinwu.qian@ua.edu}
\credit{Conceptualization, Formal analysis, Methodology, Writing - Review \& Editing, Supervision}

\affiliation[a]{Department of Civil, Construction and Environmental Engineering, The University of Alabama, Tuscaloosa, 35487, AL, U.S.}

\affiliation[b]{School of Traffic and Transportation Engineering, Tongji University, Shanghai, China}

\begin{abstract}
In this study, we show that distance-based accessibility measures of civil infrastructure may fail to aptly address the distinct challenges associated with accessing Public Charging Stations (PCSs), and we highlight the importance of incorporating the availability of opportunities near PCSs. Specifically, we distinguish between two distinct scenarios: one where charging access is secondary to opportunity access and the other where opportunity access is secondary to charging access. We conduct a comprehensive comparison between distance-based accessibility and the new opportunity-centric measures and further perform counterfactual analyses under varying PCS deployment strategies, utilizing data from over 28,000 PCSs and more than 5.5 million points of interest across twenty major metropolitan areas in the U.S. 
Our analysis reveals substantial inequalities in PCS accessibility across all metropolitan areas, where the PCS accessibility in the top 1\% of census block groups (CBGs) can be 3.6 times better than in the bottom 50\%. We also statistically confirm that an increase in PCSs will further amplify these inequalities.
In terms of opportunity-based accessibility, we show access disparities that are universal across all metropolitan areas. Importantly, inequalities stemming from the conventional accessibility definition are further exacerbated when considering opportunity availability near PCSs. Our results suggest that access inequality is linked to the income levels of CBGs, with improved opportunity-centric accessibility significantly associated with higher income in over 60\% of the metros. Despite existing disparity issues, our counterfactual analyses of different PCS deployment strategies indicate that the equity-centric deployment strategy based on distance measures turns out to be the least equitable when opportunities are considered. Instead, the current PCS deployment is found to achieve a higher POI attachment and Charging-induced Activity Accessibility score across all 20 metros over alternative strategies, with a median increase in scores by at least 63\%. Nevertheless, the opportunity-centric performance of PCS deployments remains to be largely affected by the pre-existing unequal distribution of opportunities in our cities. Our results underscore the complexity of locating charging stations and provide important implications that may guide the design of nationwide charging networks for long-term societal benefits.
\end{abstract}


\maketitle

\section{Introduction}\label{section:intro}
Reflecting a significant shift towards electrified transportation, global Electric Vehicle (EV) sales have surged tenfold, from 1 million in 2017 to 10 million in 2022~\cite{ieaGlobalOutlook}. Accompanying the growth of EVs, the rapid development of public charging stations (PCSs) plays an essential role in supporting this transition. In the U.S., motivated by the pursuit of achieving net zero emissions by 2050, states across the nation, in conjunction with the federal government, are rapidly advancing toward the implementation of EV charging stations to reach the target of 500,000 EV chargers by 2030~\cite{dotHistoricStep, whitehouseFACTSHEET}. Beyond its obvious necessity for those without home charging capabilities, the public charging infrastructure is equally vital for those with home charging, providing essential top-ups and offering a safety net for longer journeys. Recent studies suggest that by 2030, approximately 53\% of EV owners will have access to home charging~\cite{pierce2023home}, leaving a significant portion reliant solely on public infrastructure, and PCSs are projected to serve at least 30\% of all charging needs. Consequently, developing an extensive, accessible, and efficient public charging network is integral to sustaining and further accelerating the global EV boom.
Despite the strong momentum in EV penetration and PCS expansion, one essential research question remains open: How do we evaluate the effectiveness and efficiency of our charging station deployment?

Providing an answer to the above question is critical, as it will guide policy design that can lead to long-term community benefits. Among various metrics that could describe the multifaceted performances of infrastructure, accessibility remains a central factor. It quantifies how effectively an infrastructure plan serves the community, based on its definition ``\textit{the ease and convenience of access to spatially distributed opportunities with a choice of travel}''~\cite{dong2006moving}. 
Initial methods of measurement, such as distance or time to the nearest service, population-to-provider ratios, and cumulative opportunity measures, offer ease of understanding and interpretation. However, they tend to oversimplify the relationship between supply and demand regarding accessibility and fail to account for the effects of competition for the available opportunities~\cite{luo2012variable, xing2020environmental, li2022subway}. To counter these shortcomings, the class of gravity models was developed, taking into account factors like proximity, availability, and competition. While these models marked an improvement, they have attracted criticism for their complexity, interpretive challenges, and an over-reliance on the selection or empirical determination of the distance-decay function~\cite{mcgrail2009measuring, yang2023understanding}. 
Besides these known shortcomings, we further argue that the current definition of accessibility is too generic and lacks specific adaption to fully capture the distinct attributes of charging events. To understand this, we must acknowledge the fact that, unlikely conventional travel activities, charging does not necessarily constitute a primary travel decision and, in many cases, are secondary to daily activities. To this end, we claim that the evaluation of accessibility should not only cover the physical hurdle in reaching a particular station but also embed the opportunity of engaging in activities around the charging locations. 

This paper initiates an exploration into extending the definition of accessibility to PCS. It introduces new dimensions that capture the unique characteristics of charging events and provides a quantitative and qualitative assessment of the performance of the existing charging infrastructure in ten major U.S. metropolitan areas. We will explore how traditional accessibility measures present PCS performance. Moreover, our analysis will go further by connecting accessibility with increased opportunity access. We will deeply investigate two novel measures: one focusing on the convenience of charging a vehicle while simultaneously engaging in other activities - the primary emphasis here being on the activities themselves with a proximal PCS serving as a secondary benefit; the other scenario underscores the direct access to PCSs, with the supplemental advantage of conducting other activities during the charging process. Our analyses leverage a dataset of over 5.5 million points of interest linked to more than 28,000 PCSs across twenty metropolitan areas in the U.S., aiming to address the following primary research questions:
\begin{enumerate}
    \item How accurate do traditional distance-based accessibility metrics characterize the performances of the public charging infrastructure?
    \item Which additional perspectives related to charging behavior should be considered to guide the evaluation and deployment of charging facilities, and in what ways do traditional methods fail to adequately address these considerations?
    \item How can we develop new accessibility metrics inspired by charging behavior for a quantitative evaluation of charging facility deployment performances, and what additional insights do these metrics offer that extend the conventional understanding of accessibility?
    \item How can we employ counterfactual analyses for a quantitative evaluation of the U.S. deployment of charging stations, leveraging both the existing and newly proposed metrics?
\end{enumerate}

By examining these research questions, our study aims to highlight the complexity in guiding and evaluating the deployment of charging stations, provide valuable insights into the factors shaping the EV charging landscape, guide the development of effective policies, and facilitate the future growth of charging infrastructure in diverse urban contexts. The rest of this study is organized as follows. First, in~\Cref{section:background}, we provide an overview of public charging as a chance to participate in nearby activities and review the strategies to deploy and examine the charging infrastructure accessibility. ~\Cref{section:data} presents the main data collection and processing steps. Next,~\Cref{section:method} outlines gravity and opportunity-centric metrics for accessibilty to charging stations. We then provide the results of our comprehensive analysis in~\Cref{section:results}. Finally,~\Cref{section:discussion} provides a discussion by summarizing our key findings, conclusions, and future direction.
\section{Background}\label{section:background}
\subsubsection*{Deployment of Public Charging Stations} 
EV charging infrastructure plays a crucial role in the U.S.'s ambitious efforts to combat climate change, mainly by removing a main long-term challenge and creating a positive loop that enforces EV adoption~\cite{lee2018charging}. Currently, home location chargers are the preferred charging venue, where more than 50\% of charging sessions occur~\cite{hardman2018review}. Followed by work location charging, public chargers remain indispensable for EV users who lack access to home charging or face limitations in this regard~\cite{hardman2018review}. This is particularly the case for residents in densely populated urban areas, where off-street parking and private garages are less accessible and the installation cost of chargers in multi-unit dwellings (MUDs) is prohibitively high~\cite{ge2021there, zhang2023electric}. 

At the same time, EV sales in the US experienced an increase of 55\% relative to 2021 due to a plethora of factors such as purchase subsidies and an expanded range of available models~\cite{ieaGlobalOutlook}. However, alongside this surge in EV demand, developing a public charging infrastructure still requires addressing a multi-dimensional challenge to ensure that the infrastructure can adequately meet the escalating user needs. Recent studies have mainly explored the charging station location problem, taking into account traffic flow dynamics and drive range~\cite{chen2020optimal, bao2021optimal,qian2019stationary}, charging cost~\cite{guo2020impacts}, waiting times~\cite{uslu2021location}, electric power distribution loads~\cite{saadati2022effect}, or a combination of such factors~\cite{loni2023data}. Furthermore, the development of public charging infrastructure must prioritize the equitable distribution of costs and benefits throughout society. Achieving this goal requires acknowledging existing equity gaps and the limitations in our current methods of evaluating the performance of the existing charging layout. This understanding is imperative for addressing disparities and ensuring that the infrastructure serves the needs of all segments of society effectively.

\subsubsection*{Public Charging Stations as an Opportunity for Activities}
The significant uptake of EV sales has resulted in a growing body of research focusing on EV drivers' charging preferences. Studies have found a plethora of factors including charging duration~\cite{bruckmann2023experimental, andrenacci2021modelling, globisch2019consumer}, the need for detours~\cite{liao2017consumer, guo2022modeling}, and range anxiety~\cite{jia2023investigating, ardeshiri2020willingness} play a vital role in the selection of charging stations. Given that charging can take several hours depending on the vehicle's battery size and charging power—for example, charging a Tesla Model 3's battery from 40\% to 50\% on a Level 2 charger requires 30 minutes, and reaching to 90\% takes an additional 90 minutes~\cite{blinkcharging}—drivers often look for ways to occupy this time. This prompts users to engage in auxiliary activities, particularly when options are proximate in a walking distance. Lei et al. found that instead of visiting the nearest charging station, 40\% of drivers are willing to drive an extra 1.5 kilometers to reach a station that better suits their additional needs, like the presence of amenities including rest areas and dining options~\cite{lei2022understanding}. Simultaneously, the growing link between businesses and charging stations is the other factor enforcing the attachment between the charging behavior and the activities. Recent studies highlight that over 50\% of public charging stations are strategically positioned near commercial venues, where the pricing often deviates from the optimal pricing strategies~\cite{arlt2023retail}, followed by a captured trend where nearby businesses experience a 4\% increase in monthly visits after the installation of Tesla Superchargers in U.S.~\cite{babar2023recharging}. 

\subsubsection*{Equity Implications of Existing Public Charging Infrastructure}
Recent studies have identified a strong correlation between the placement of public EV charger installations and the ownership of electric vehicles~\cite{hsu2021public, wells2012converging}. However, while rational for the early stages of charging infrastructure development, such an approach can also lead to or amplify the existing socioeconomic inequities among population groups. Consequently, as EVs become more affordable and offer longer driving ranges, the layout of public charging infrastructure can have significant and lasting implications for determining which segments of the population can benefit from the advantages of EV adoption~\cite{carley2019evolution}. In this regard, several studies have investigated the socioeconomic equity concerns surrounding the distribution of charging stations within communities~\cite{hsu2021public, nazari2022toward, khan2022inequitable, roy2022examining, li2022spatial, liang2023effects}. For instance, a study conducted in California examined the presence of charging stations in different census block groups and revealed disparities in public charging access, particularly in lower-income block groups~\cite{hsu2021public}. The research also highlighted that areas with a higher concentration of multi-unit housing, where residential charger access is less common, experienced more pronounced disparities in public charger availability. Current data also reveals the local benefits associated with offering more charging opportunities to be unequally distributed. For instance, using 14 million housing transactions for almost three decades showed that charging infrastructure can be capitalized into property values where the average price premium for houses with PCSs within 0.5 km compared with houses without proximate PCSs~\cite{liang2023effects}. Others have highlighted the potential ramifications of inadequate charging infrastructure support among communities, leading to missed environmental and health benefits at both the household and community levels. These include reduced local air pollution, cost savings compared to traditional internal combustion engine vehicles, and advantages of leveraging regional and local power grid ancillary services, including localized voltage support~\cite{pan2019potential, palmer2018total, knezovic2016enhancing}.

\subsubsection*{Performance Measures in Deployment of Public Charging Stations}
Despite the growing evidence of disparities in charging station deployment, the current approach to equity evaluation for charging infrastructure often mirrors the spatial analysis and conventional accessibility metrics used for other civil infrastructures~\cite{li2022spatial, guo2021disparities, hsu2021public, carlton2022electric}. For instance, users' accessibility to charging infrastructure is measured by counting the available charging opportunities within a 15-minute driving range~\cite{carlton2022electric}. Other methods for equity assessment of charging layout include consideration of distance from the charging demand point to the stations~\cite{xu2022optimal}, travel demands and traffic flow volume passing through each region~\cite{do2018equitable}, determining a threshold for the distance (or travel time) between the users and the charging stations \cite{iravani2022multicriteria}, and incorporating binary variable representing the presence of at least one EV charging station in a particular zip code and The total number of EV charging stations in each zip code before performing a correlation analysis with socioeconomic factors \cite{khan2022inequitable}. Nevertheless, while they might still be appropriate in other types of civil services like urban public facilities, such approaches fall short of considering the intricate urban dynamics involved in users' charging station selection. This complexity arises in part from the need to consider the spatial arrangement of various locations and their interaction with the unique characteristics of neighborhoods and road networks within the local area and in part due to the users' charging behaviors that are closely entwined with their day-to-day activities~\cite{toth2021inequality,lee2018charging}. Furthermore, examining the deployment of PCSs requires special attention to the interplay of factors such as installation incentives and policies, user behaviors, and the underlying business strategies driving the deployment process, which can vary greatly in different regions. 
\section{Data}\label{section:data}
\subsection{Data Collection}
In this study, multi-source data capturing geolocation of PCSs, sociodemographic information at the census block group level (CBG), and the spatial distribution of major points of interest (POIs) are collected and analyzed to gain insights into opportunity-centric PCS accessibility within 20 U.S. major metropolitan areas. 

\noindent \textbf{AFDC Public Charging Station Locator:} The identification of PCSs is based on the DOE's Alternative Fuels Data Center (AFDC) as the most comprehensive charging station locator in the U.S.~\cite{energyEEREAlternative}. The dataset includes over 65,000 public charging stations and more than 170,000 Electric Vehicle Supply Equipment (EVSE) across the U.S. as of January 2024. 

\noindent \textbf{American Community Survey data:} Sociodemographic data for the communities are sourced from the 2016-2020 American Community Survey (ACS) 5-year estimates, providing detailed information on population and income status at the census block group (CBG) level~\cite{censusAmericanCommunity}. The dataset includes over 242,000 CBG across the U.S., with a total population exceeding 329 million.

\noindent \textbf{SafeGraph Points-of-Interest data:} To gather information on the location and activity categories of various places, we leverage the SafeGraph Global Points of Interest (POIs) data~\cite{safegraphGlobalPoints}. The dataset comprises information on more than 18.5 million unique places spanning across the United States as of January 2024. Specifically, along with the geometric coordinates, the dataset includes brand affiliation and two levels of category tags for each place. Additionally, each place is classified according to the North American Industry Classification System (NAICS) codes, encompassing 607 distinct NAICS codes and 253 major category tags such as `Restaurants and Other Eating Places' and `Offices of Physicians'. Listed POIs are evaluated to be no more than 10 meters away from their location on Google Maps, where accuracy for top brands' attributes is reported as high as 99.9\% compared to brand store locators~\cite{safegraph_pois}.

\subsection{Data Processing}
\subsubsection*{Selection of Metro Areas}
We narrow our analysis of the U.S. charging infrastructure by focusing on major metropolitan areas rather than state or regional levels. This decision is based on two reasons. Firstly, a noticeably higher rate of EV adoption and deployment of EV infrastructure is evident in larger metropolitan areas compared to smaller urban or rural regions. This distinction allows for a more meaningful analysis of accessibility and the interaction between activities and charging facilities located in close proximity to each other. Second, even in CBG resolution, rural areas have larger geographic units with considerably less road network coverage. This can impact the accuracy of our accessibility analysis, as a charging station located on one side of a block group may not be accessible to the other side. 

\begin{figure}[h]
	\centering
	\includegraphics[width=0.8\textwidth]{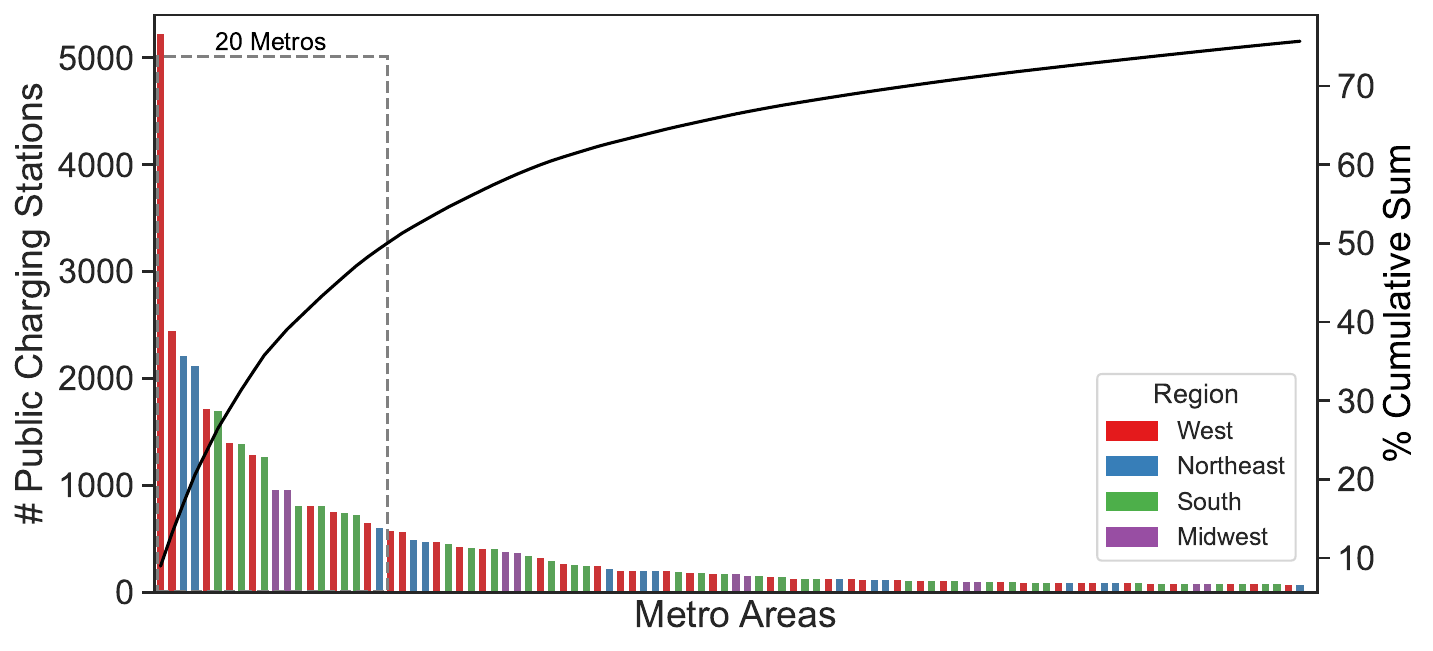}
	\caption{Number of Public Charging Stations in U.S. Metro Areas}
	\label{fig:metros_pcs_count}
\end{figure}

\begin{figure}[h]
	\centering
	\includegraphics[width=0.8\textwidth]{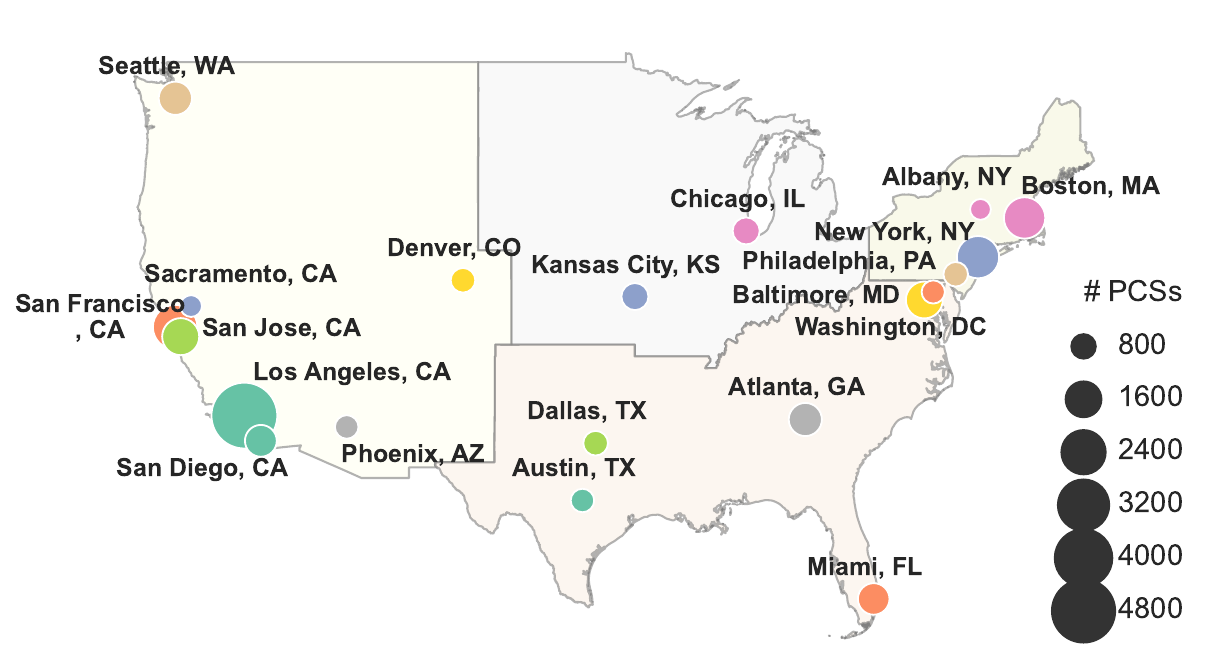}
	\caption{Selected U.S. Metro Areas}
	\label{fig:metros}
\end{figure}

Next, to determine a representative sample of metro regions across the U.S., we examined the distribution of PCSs in metropolitan areas, as depicted in~\Cref{fig:metros_pcs_count}. \Cref{fig:metros_pcs_count} reveals that the top 20 metropolitan areas, in terms of PCS count, collectively represent 49.3\% of the total PCSs located in metro areas- that is, 28,762 out of 58,321 PCSs. Therefore, our further analysis focuses on these twenty major metro areas in the U.S., boasting an extensive charging infrastructure network that is distributed across different regions. \Cref{fig:metros} displays the selected metro areas. Among the selected metros (and nationally), the Los Angeles metro area has the largest charging infrastructure, offering 4,232 public charging stations and 13,957 EVSEs. In contrast, the Albany metro area has the smallest infrastructure within the selected metros, with 611 PCSs and 1,676 EVSEs.

\FloatBarrier

\FloatBarrier

\subsubsection*{Selection of Activities}
To analyze the opportunity-centric accessibility to charging infrastructure, we curated the SafeGraph POIs dataset by applying multiple essential filters to ensure relevance to charging station activity. Initially, we excluded specific business categories not directly related to charging. This led to the elimination of seven high-level categories as defined by the North American Industry Classification System (NAICS)~\cite{naics2022}, including `Administrative and Support and Waste Management and Remediation Services,' `Agriculture, Forestry, Fishing and Hunting,' `Mining, Quarrying, and Oil and Gas Extraction,' `Management of Companies and Enterprises,' `Utilities,' `Construction,' and `Manufacturing.' These categories collectively represented 7.5\% of the total POIs. Further refinement was achieved by examining the median dwell time at various establishments using Advan's spend patterns dataset at individual POIs~\cite{safegraphSpend}. We excluded any location where the median dwell time classified by their NAICS code was over 5 hours or less than 10 minutes. This criterion led to the removal of places under category tags such as `Financial Transactions Processing, Reserve, and Clearinghouse Activities' and `Petroleum Bulk Stations and Terminals'. Moreover, we identified and excluded POIs that, despite a high count, did not facilitate engagement with charging activities. This includes ATMs and specific in-store services like Redbox vending machines within larger retail outlets, as well as POIs such as USPS Collection Points, UPS Drop Boxes, Allpoint ATMs, and Redbox locations. Finally, we filtered out locations designated as alternative fuel stations, including sub-category of `Other Gasoline Stations' and specific brands such as `ChargePoint Network Charging Station', `EVgo', `Tesla Destination Charger', and `Blink Charging Co.', to avoid redundancy with PCS data.

After this comprehensive filtering process, we retained 16,544,048 POIs, which constitute 86.3\% of our original dataset. The selected POIs were then classified into nine groups based on their NAICS high-level business categories, as follows:
\begin{itemize}
    \item Accommodation and Food Services (e.g., Restaurants \& Other Eating Places, Drinking Places - brands like Subway, and Starbucks)
    \item Retail and Wholesale Trade (e.g., Supermarkets, Convenience Stores, Sporting Goods Stores - brands such as CVS, and 7-Eleven)
    \item Health Care and Social Assistance (e.g., Offices of Physicians, Offices of Dentists, Child Day Care Services - brands such as DaVita, and Aurora Healthcare)
    \item Arts, Entertainment, and Recreation (e.g., Fitness \& Recreational Sports Centers, Golf Courses \& Country Clubs, Casinos - brands like CrossFit, Club Pilates)
    \item Professional, Scientific, and Technical and Educational Services (e.g., Offices of Lawyers, Elementary \& Secondary Schools - organizations like H\&R Block, and Mathnasium)
    \item Public Administration and Other Services (e.g., Religious Organizations, Public Offices, Drycleaning \& Laundry Services - brands like Rotary Club)
    \item Real Estate and Rental and Leasing (e.g., Lessors of Real Estate, Malls - brands like Enterprise Rent-A-Car)
    \item Finance, Insurance, and Information (e.g., Insurance Carriers, Wireless Telecommunications Carriers, Libraries \& Archives - brands like Edward Jones, T-Mobile) 
    \item Transportation and Warehousing (e.g., Used Household \& Office Goods Moving, General Warehousing \& Storages - brands like ArcBest, iStorage)
\end{itemize}

Finally, the datasets were mapped onto the census block group (CBG) level within the major metro areas by using geometric boundaries from the 2020 TIGER urban shapefiles~\cite{censusTIGERLineShapefiles}. We note that our definition of metro areas aligns with the U.S. Census delineations of a Metropolitan Statistical Area (MSA)~\cite{ratcliffe2016defining}. For instance, the New York, NY metro area corresponds to New York--Jersey City--Newark, NY--NJ, and the Los Angeles, CA metro area corresponds to Los Angeles--Long Beach--Anaheim, CA. 
\section{Method}\label{section:method}
We describe several metrics to evaluate the effectiveness of public charging station layouts, incorporating traditional accessibility aspects along with opportunity-centric measures for CBGs. We then proceed to introduce our counterfactual analysis and the alternative charging infrastructure deployment scenarios.

\subsection{Metrics}\label{section:metrics}

\subsubsection{Accessibility Measures}
Our first metric on the CBGs' accessibility to the public charging stations is based on the gravity model of measuring accessibility, a widely utilized approach in the urban accessibility literature~\cite{guida2020urban, levinson2020towards}. In our case, gravity models enable us to quantify the degree of CBGs' accessibility to public charging stations within their catchment area (defined as the nearest 20 stations from their centroid). For brevity, \textit{we will use CBG accessibility or gravity access to refer to the gravity model-based accessibility measure}, with the mathematical expression below:

\begin{align}
A_i &= \sum_{j\in S} \mathbb{I}_{ij} \times N_j \times e^{-\beta t_{ij}}, \quad \forall\,i\in Z
\label{eq:cbg_accessibility}
\end{align}

where $A_{i}$ represents the accessibility score of a CBG $i$, $\mathbb{I}_{ij}$ is an indicator variable that takes the value of 1 if station $j \in S$ is within the catchment area (nearest 20 stations) of CBG $i$, and 0 otherwise. $N_j$ denotes the number of EVSEs available at charging station $j$, and $t_{ij}$ indicates the distance (km) between the centroid of CBG $i$ and charging station $j$. The impedance factor $\beta$ influences the rate at which the accessibility score decreases with increasing distance and is set to $\beta=0.08$~\cite{levinson2020towards}. 

\subsubsection{Opportunity Score and Opportunity-centric Accessibility Measures}
The gravity-based accessibility measures characterize spatial impedance to and from a charging station but fall short of considering EV users' access to additional opportunities while charging their vehicles. In many cases, users are willing to sacrifice travel time in pursuit of engaging in additional activities around the charging station. This asserts the need to examine the number of points of interest (POIs) nearby a charging station as a proxy for additional activity opportunities. 

Thus, we begin by first introducing the POI opportunity score of a station $j$ for POI category $c$, $P_j^c$, written as:

\begin{equation}
    P_j^c= \sum_{k\in K^c} \frac{\mathbb{I}_{kj}}{|K^c|}
\end{equation}
where $K^c$ is the complete set of POIs in category $c$, $\mathbb{I}_{kj}$ is an indicator variable and takes the value of 1 if a POI $k$ is associated with (within 250 meters from) the charging station $j$ and 0 otherwise. We note that this measure is scale-invariant by diving the size of the set $K^c$, so that the measure is comparable across different metro areas despite the differences in the total number of POIs. $P_j^c$, in this case, provides additional opportunity insights beyond traditional accessibility measures, which will shed light on if certain groups of activities are (disproportionally) concentrated near charging stations and allow us to examine the disparities in accessing opportunities across all PCSs.  

Based on the POI score, we extend the conventional accessibility measure by concurrently examining the physical impedance to reach a PCS and the opportunity cost when charging at the PCS. Nevertheless, we need to acknowledge the difference between two distinct scenarios:  (1) an EV user visiting a place as the primary decision and charging at that location as a derivative activity, and (2) an EV user visiting a charging station as the primary decision and engaging in extra activities as the derivative decision. This gives rise to two distinct opportunity-centric accessibility measures as detailed below:

\textbf{Activity-Induced Charging Accessibility (AICA)}: This metric corresponds to scenario (1) and measures how PCSs are easily accessible/available when EV users engage in their daily activities from their home locations (the centroid of CBG in our case). For example, EV users can utilize associated charging stations at their nearby shopping venues. Mathematically, we express the AICA of a CBG $i$ as:

\begin{align}
AICA_{i}^{c} &= \sum_{k\in K^c_i} \sum_{j\in S} \mathbb{I}_{kj} \times N_j \times e^{-\beta t_{ik}}
\label{eq:aic}
\end{align}
where $K^c_i$ is the set of the nearest 20 POIs in category $c$ to CBG $i$ and $\mathbb{I}_{kj}$ is the indicator variable that takes the value of 1 if a PCS $j$ is within 250 meters of POI $k$ and 0 otherwise. $AICA_i^c$ agglomerates both physical impedance from a CBG $i$ to nearby activity locations $k\in K_i^c$ and the availability of accessing a charging port at PCSs associated with each POI. 

\textbf{Charging Induced Activity Accessibility (CIAA):} This metric corresponds to scenario (2) and measures the number of opportunities in category $c$ when EV users from CBG $i$ access PCSs in its catchment area (nearest 20). Mathematically, we express CIAA as:

\begin{align}
CIAA_{i}^{c} &= \sum_{j\in S}\sum_{k\in K^c} \mathbb{I}_{ij} \times \mathbb{I}_{kj} \times e^{-\beta t_{ij}}
\label{eq:cia}
\end{align}
where $CIAA_i^c$ combines the physical impedance from CBG $i$ to the nearest 20 PCSs and accumulates the number of POIs in category $c$ associated with each PCS. As a consequence, this metric considers the scenario when an EV user selects to charge at PCS $j$ and gauges the availability of engaging in additional activities rather than waiting in the vehicle.

\subsection{Counterfactual Analyses}\label{section:scenarios}
Merely quantifying the accessibility score, regardless of whether considering proximate opportunities or not, does not paint a complete picture that enables the equality assessment of a PCS deployment plan. This must be supported by comparing the current deployment plan with feasible alternatives so that both quantitative and qualitative assessments can be made. In this regard, we conduct counterfactual analyses to compare the existing PCS layout in major metro areas with four alternative scenarios representing four different planning philosophies using the above metrics. The four alternative scenarios are as follows:
\begin{enumerate}
  \item \textbf{Uniform}: Under the uniform strategy, current PCSs are reassigned spatially uniformly across the metro area. This approach allowed us to establish a baseline for comparison, providing insights into the distribution of stations when no particular factors other than a uniform spatial coverage were taken into account.
  \item \textbf{Population-based}: In the population-based strategy, current PCSs are redistributed proportional to the population density of CBGs, with more PCSs in CBGs with higher population density. 
  \item \textbf{Profit-based}: The profit-based strategy involves reassigning existing PCSs based on the combination of median-household income and the population of CBGs. By examining this scenario, we seek to investigate the potential disparities that may arise from prioritizing economically advantaged areas.
  \item \textbf{Equity-based}: The equity-based strategy reallocates PCSs to make accessibility to PCSs equitable across all CBGs. This approach aims to understand the gains and losses of equity-driven charging station deployment that may arise because of additional opportunity considerations. The equity-based reallocate is achieved by solving the integer linear assignment problem with the objective function being $\text{Max} \,  \text{Min}_{i\in Z} A_i$ and is subject to the total number of PCSs in the current plan. 
\end{enumerate}

In our counterfactual analysis, we make two adjustments to ensure the accessibility outcomes across various scenarios are directly comparable to the current configuration of charging infrastructure. Firstly, acknowledging that the number of EVSEs can be different across charging stations, we equate the number of EVSEs at all stations across the scenarios and the existing layout, allocating one EVSE per station. Secondly, to have a realistic comparison of distance and nearby opportunities between the scenarios that aligns with the existing layout, we reposition the designated stations to the closest road node within the transportation network.
\section{Results}\label{section:results}
In line with our research questions, we present our results in the following order: First, we examine CBGs' accessibility within and between the metro areas, focusing on gravity-based measures. Subsequently, we extend our analysis to include the opportunity-centric metrics introduced in this study, namely Charging-Induced Activity Accessibility (CIAA) and Activity-Induced Charging Accessibility (AICA). Following this, we explore the association between CBGs' accessibility and income levels, highlighting the income-based disparities in access to charging infrastructure within large U.S. metropolitan areas. Concluding our results, we perform a counterfactual analysis to evaluate the efficacy of the existing PCS deployment layout against four alternative strategies, providing comprehensive insights into the implications of different deployment approaches.

\subsection{Accessibility to Public Charging Stations}\label{section:results_gravity_access}
\subsubsection*{Disparities in PCS Accessibility through Gravity-based Approach}
\begin{figure}
	\centering
	\includegraphics[width=0.95\textwidth]{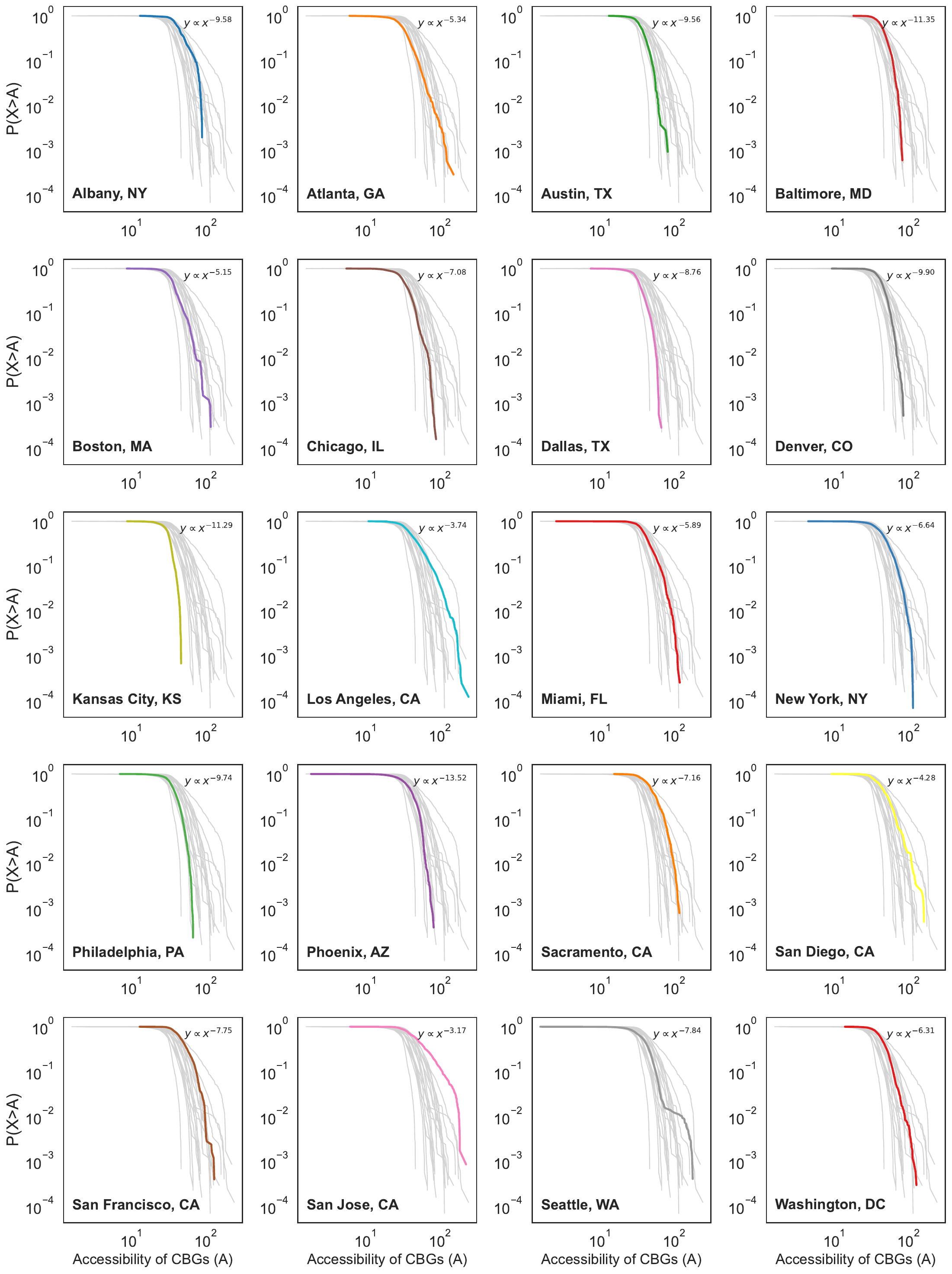}
	\caption{Census Block Groups' Accessibility to Public Charging Stations}
	\label{fig:cbg_access_income}
\end{figure}

We first present the results regarding the CBGs' gravity-based accessibility to PCSs in~\Cref{fig:cbg_access_income}. The figure visualizes the complementary cumulative density function (CCDF) for the distribution of PCS accessibility across twenty metro areas in a log-log scale, where each line visualizes the proportion of CBGs, $P(x>A)$, that has an accessibility score greater than $A$. As a consequence, the further each line stretches towards the right, the greater the variance of accessibility across all CBGs. 
We observe a notable difference in the accessibility of CBGs to PCSs both within and between metro areas. For each metro area, we observe that the tail of the accessibility distribution (top 10 percentile of CBGs' accessibility scores) can be well approximated by the power-law distribution, $y\propto x^{-\gamma}$, with $\gamma$ in the range of 3.17 (San Jose, CA) to 13.52 (Phoenix, AZ) for the selected metro areas. The estimated power-law exponents, $\gamma$, characterizing the tail of the accessibility distributions, are depicted as part of ~\Cref{fig:cbg_access_income}. The power-law approximation signifies that a very small proportion of CBGs have a high gravity-based accessibility score to PCSs, whereas the majority of the CBGs have poor access to the PCSs~\cite{clauset2009power}. For instance, in San Jose, CA, the top 1\% of CBGs boast a gravity-based accessibility score that is over 3.55 times greater than 50\% of the CBGs.

\subsubsection*{Charging Infrastructure Expansion and PCS Accessibility Disparities}
We are further interested in checking whether the expansion of public charging infrastructure is associated with a more gradual decline in the approximated power-law distributions, thereby exacerbating the disparity in PCS accessibility across CBGs. To explore this, we perform an Ordinary Least Squares (OLS) regression, using the power-law exponent parameter ($\gamma$) as the dependent variable and the number of PCSs (X) as the independent variable. The results suggest that $ln(\gamma)\sim -0.49 ln(X)$, with p-value=$0.001$ and R-squared value=$0.49$. This indicates that a 1\% increase in the number of public charging stations will result in 0.49\% reduction in the power law coefficient $\gamma$. This statistically significant relationship, confirmed at the 0.05 significance level, leads us to conclude that PCS network growth tends to diminish the rate at which the CCDF of accessibility scores decays, hence resulting in greater inequality for the PCS accessibility among all CBGs. This is while CBGs in metros with a higher number of stations do not hold a higher accessibility score on average (for the top 10 metros, that is 14.8\% lower than those of the bottom 10).

\FloatBarrier
\FloatBarrier

\subsection{Distribution of Opportunities around Public Charging Stations}
\subsubsection*{Activities Attachment to Charging Infrastructure}
Before delving into the opportunity-centric accessibility measures, it is crucial to map out the extent to which POIs are linked to PCSs across various place types and metropolitan areas. Thus, our initial analysis focuses on the distribution of over 5.5 million selected POIs within metro areas, finding that 1.4 million of these are located within a 250-meter radius of a PCS. In~\Cref{fig:ccdf_poi_places_a}, we assess the spatial arrangement of POIs around PCSs in comparison to their general distribution throughout the metro area. We further categorize the POIs into two groups: `elastic' activities, which are associated with shorter dwell times and for which alternatives are readily available (like dining places or fitness centers), versus `inelastic' activities, characterized by longer dwell times and fewer alternatives (such as churches, government offices, and workplaces).

We report that there is a notable difference in the presence of certain POI categories near PCSs as opposed to their distribution metro-wide, with a distinct preference for PCSs to be located near sites of elastic activities. Specifically, categories such as ``Accommodation \& Food Services," ``Finance, Insurance, and Information," and ``Professional, Scientific, Technical, and Educational Services" are more densely populated around PCSs, exceeding the metro-wide percentage for these categories by 22.2\%, 26.0\%, and 39.2\%, respectively. Within the ``Accommodation and Food Services" category alone, the variation spans from 7.2\% for Snack and Nonalcoholic Beverage Bars to 58.2\% for Casinos. Conversely, inelastic activities like ``Public Administration and Other Services" and ``Transportation and Warehousing" show a 19.8\% and 39.2\% reduction in presence near PCSs, respectively. These results suggest that PCSs are more commonly found near locations conducive to shorter, flexible activities, offering users a variety of options to utilize their time effectively while charging. However, those seeking PCSs near their less flexible, inelastic activity locations may find fewer options available, potentially narrowing the range of activities they can engage in during charging sessions.

\subsubsection*{Equality of the Number of Opportunities at Charging Stations}
\begin{figure}
  \centering
  \begin{subfigure}[b]{0.325\linewidth}
    \centering
    \includegraphics[width=\linewidth]{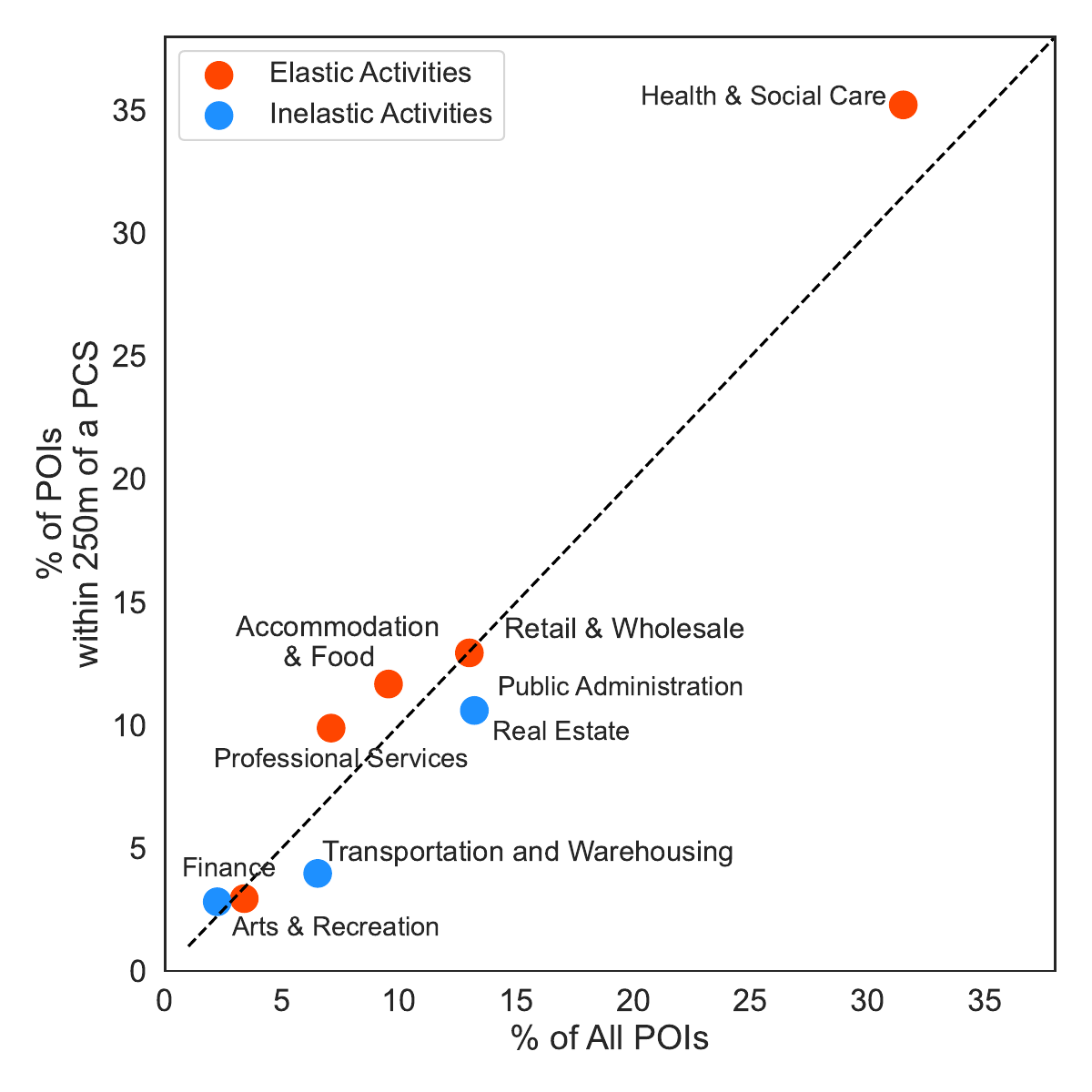}
    \caption{Spatial Arrangement of POIs}
    \label{fig:ccdf_poi_places_a}
  \end{subfigure}
  \hfill
  \begin{subfigure}[b]{0.666\linewidth}
    \centering
    \includegraphics[width=\linewidth]{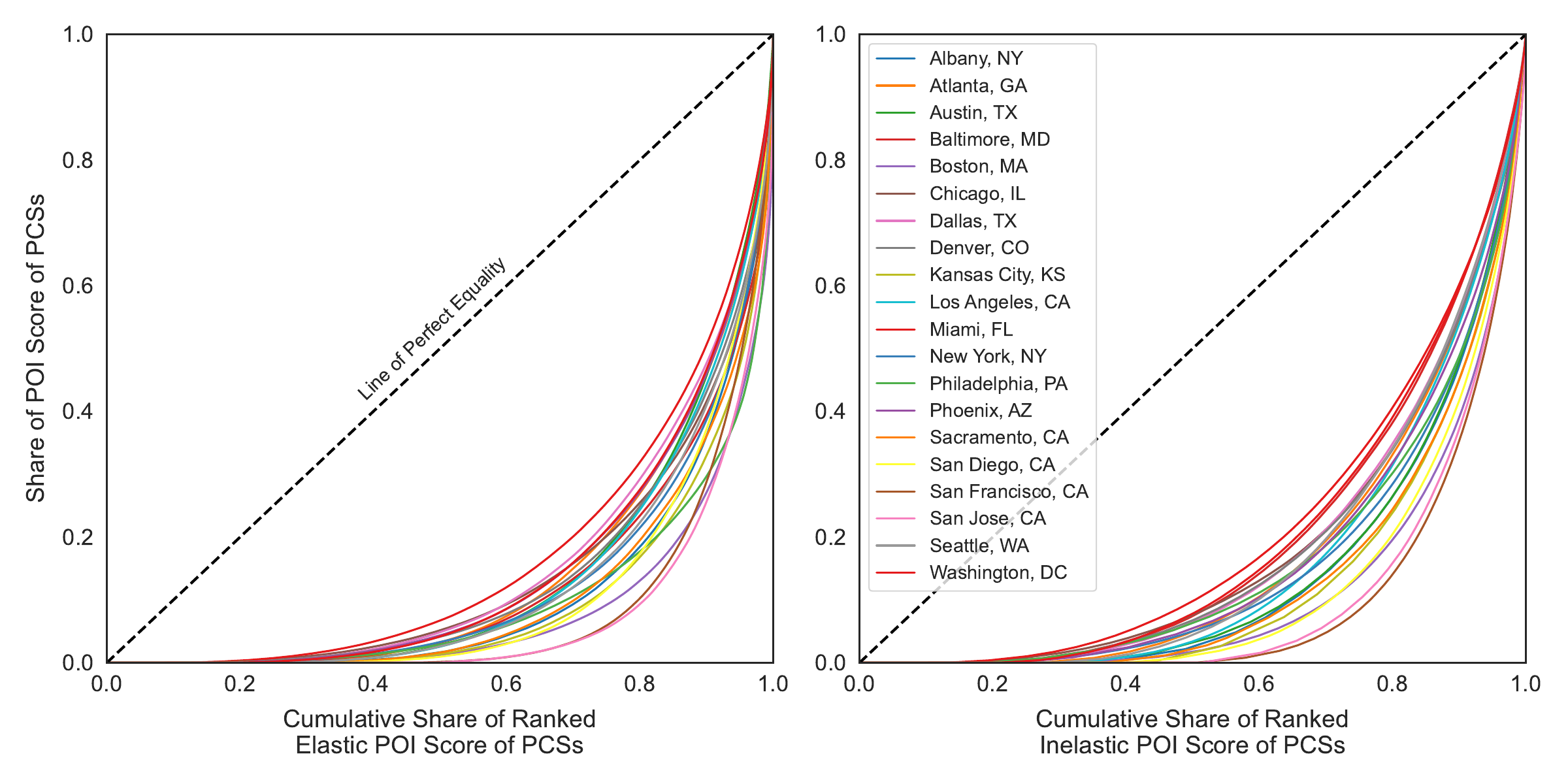}
    \caption{Lorenz curve of POIs Associated with Public Charging Stations}
    \label{fig:ccdf_poi_places_b}
  \end{subfigure}
  \caption{POIs Associated with Public Charging Stations}
  \label{fig:ccdf_poi_places}
\end{figure}

Besides the differences in overall POI concentration, we also observe a significant inequality in the associated number of POIs across all the PCSs, as depicted by the Lorenz curves in~\Cref{fig:ccdf_poi_places_b}. The outcomes highlight a pronounced inequality in the availability of opportunities at various PCSs, a trend consistent across metropolitan areas for both categories of activities, elastic and inelastic. In all 20 metro areas, a few numbers of PCSs are associated with a very high number of POIs (more than 50\% of the cumulative share of POI scores corresponds to less than 10\% of the PCSs). We can further quantify the degree of inequality via Gini index~\cite{gastwirth1972estimation}, calculated as the distance of each curve from the diagonal line (line of perfect equality). The Gini coefficient ranges from 0 to 1, with 0 representing perfect equality and 1 denoting total inequality. The San Jose, CA metro area is found to be associated with the highest level of inequality, with Gini coefficients of 0.86 for elastic activities and 0.80 for inelastic activities. San Fransisco, CA, closely follows with Gini coefficients of 0.84 and 0.81 for elastic and inelastic activities, respectively. The Gini coefficients for all other cities are at least 0.59 (Miami). In conclusion, while having abundant activity locations near PCSs is desirable, the existing public charging layout sees a significant imbalance in access to opportunities. Such disparities motivate us to take a more nuanced approach to evaluating accessibility by integrating the traditional accessibility metrics with opportunity-based measures to understand better and address the existing accessibility gaps in PCS deployment.

\FloatBarrier
\FloatBarrier

\subsection{Opportunity-centric Accessibility to Public Charging Stations}
The findings discussed thus far offer an in-depth examination of the disparities in access to PCS through the lens of a gravity-based approach. Nevertheless, we recognize that traditional accessibility measures only provide a snapshot of the performances of PCS infrastructure. Building upon this foundation, we focus on uncovering if the issues identified through conventional accessibility evaluations are mirrored or even magnified when considering the availability of opportunities associated with PCSs.

\subsubsection*{Charging-induced Activity Accessibility and Activity-induced Charging Accessibility}
\begin{figure}
  \centering
  \begin{subfigure}[b]{0.4\linewidth}
    \centering
    \includegraphics[width=\linewidth]{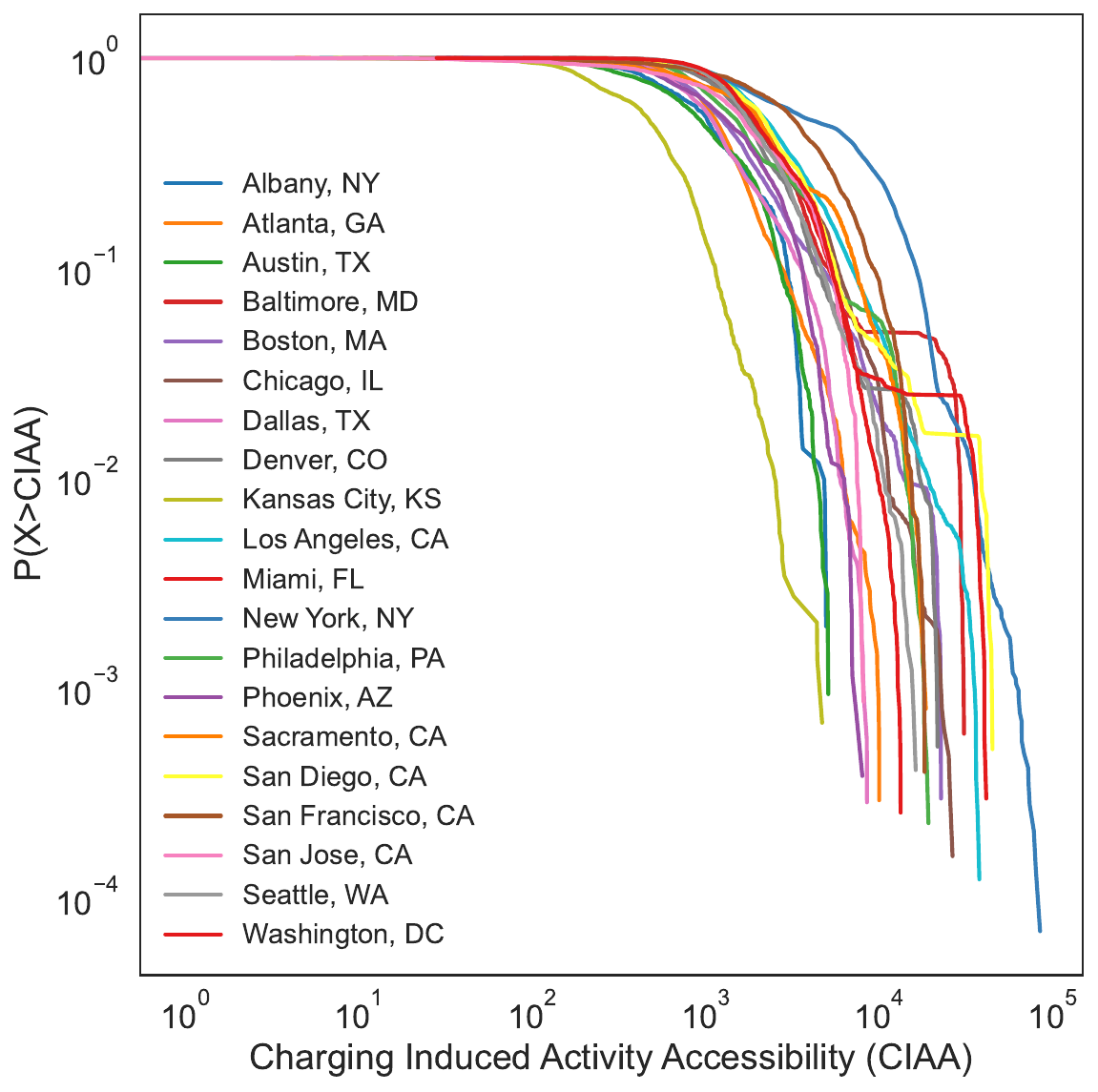}
    \caption{Charging-induced Activity Accessibility (CIAA)}
    \label{fig:aic_cia_a}
  \end{subfigure}
  \begin{subfigure}[b]{0.4\linewidth}
    \centering
    \includegraphics[width=\linewidth]{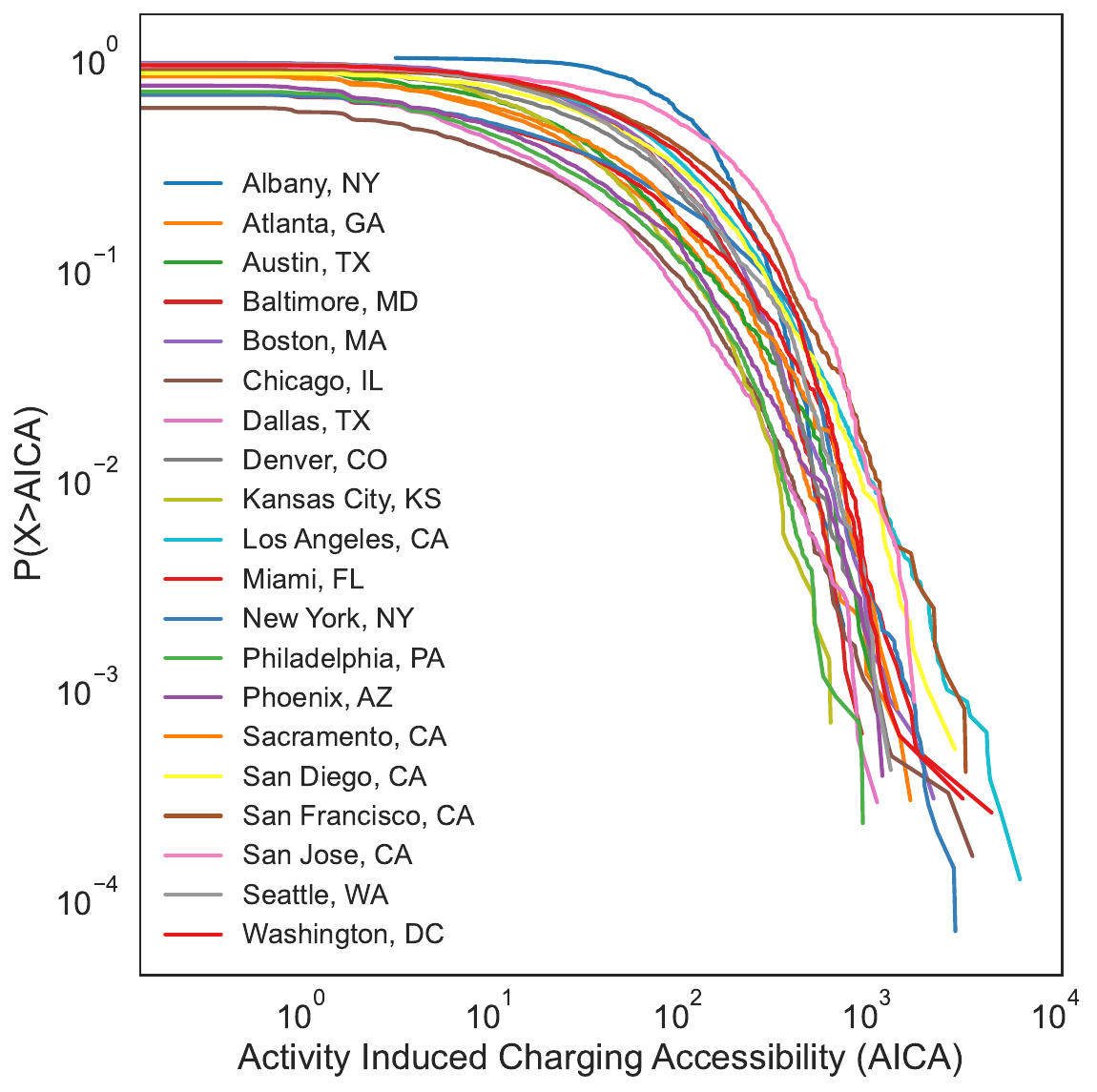}
    \caption{Activity-induced Charging Accessibility (AICA)}
    \label{fig:aic_cia_b}
  \end{subfigure}
  \caption{CCDF of CIAA and AICA Distributions Across CBGS in 20 Metro Areas}
  \label{fig:aic_cia}
\end{figure}

Similar to the results for the distribution of the accessibility scores in~\Cref{section:results_gravity_access}, we visualize the complementary cumulative density function (CCDF) for the distributions of Charging-induced Activity Accessibility (CIAA) and Actiivity-induced Charging Accessibility (AICA) in~\Cref{fig:aic_cia_a} and~\Cref{fig:aic_cia_b}. The results reveal that the tail of the CCDFs for CBGs (with top 10\% of scores) in both metrics can be well approximated by the power-law distribution, which is indicative of inequalities of opportunity-based accessibility for different communities with few CBGs having disproportionally high opportunity-based accessibility scores. The observations hold for all metro areas, and we hypothesize that such inequality issues, both in terms of conventional accessibility and opportunity-centric measures, are likely universal across major cities. Nevertheless, we observe different trends among the distributions for AICA and CIAA. This is statistically verified by fitting the power-law distributions to the CBGs with the top 10\% of the scores. The fitted results show the range of $\gamma$ between 1.30 (Austin, TX) to 3.77 (Albany, NY) for AICA and a range of 0.83 (Baltimore, MD) to 7.90 (Albany, NY) for CIAA for the fitted distributions. 

\subsubsection*{Disparities in PCS Accessibility through Opportunity-centric Measures}
We note that the $\gamma$ value range for Activity-induced Charging Accessibility (AICA) is consistently lower than that of both gravity accessibility and Charging-induced Activity Accessibility (CIAA). This suggests that the CCDFs for AICA decay slower, pointing to a broader disparity where AICA suffers a greater inequality across CBGs than CIAA. We believe this disparity is likely exacerbated due to the compounded inequality of access to PCSs and the unequal distribution of opportunities (POIs) in each metro area. Consequently, the placement of charging stations near (existing) major activity locations will inevitably lead to inequalities in opportunity-centric accessibility measures. Moreover, the $\gamma$ does not capture CBGs with low AICA scores as it focuses on the tail of the CCDF. For areas with low AICA scores, we note a significantly worse score compared to accessibility measures of other areas since there is no PCS available near their top 20 activity locations. The analysis further reveals that the disparity in AICA is more pronounced than in CIAA, as evidenced by higher $\gamma$ values for CIAA, suggesting a quicker CCDF decay and somewhat mitigated inequality. This difference underscores two primary contributors to the disparities observed with CIAA: (1) an inequitable PCS deployment that disproportionately affects CBGs, particularly those with lower incomes who may have access to fewer stations, and (2) a skewed distribution of POIs in proximity to PCSs, as indicated by opportunity scores. 

Regarding the category-wise distribution of opportunity-centric accessibility for metro areas, we observe a distinct pattern regarding access to activities. Notably, inelastic activities—those less substitutable and often essential, like government services or workplaces—tend to have higher Activity-induced Charging Accessibility (AICA) scores in 12 of the 20 metro areas studied, indicating generally better access. This is underscored by a more marked disparity in access to these inelastic activities, evidenced by lower $\gamma$ values for the top 10\% distribution in 11 metro areas when compared to elastic activities. This suggests that while access to inelastic activities is generally higher when making charging infrastructure the secondary decision after activity choice, the inequality in this access is also more pronounced. On the other hand, when examining Charging-induced Activity Accessibility (CIAA) scores, all 20 metro areas showed higher median scores for elastic activities—like dining and shopping. Moreover, in 13 metros, we observe a greater inequality in access when considering the tail of the scores with lower $\gamma$ values. This trend is the reverse of our observation with AICA, where elastic activities, being the secondary choice after deciding to charge, have higher accessibility but with a significant variation in equality of access. This analysis reveals a critical insight: focusing solely on one dimension of charging station selection and associated activities does not lead to an equitable distribution of charging access, as it underscores the intricate relationship of first and derivative travel decisions with different types of activities for EV users. 

Finally, a positive correlation between CIAA and AICA scores across all metro areas (with values as high as 0.41, 0.42, and 0.45 for New York, Miami, and Kansas City, respectively) suggests that areas with lower AICA scores also tend to exhibit poorer CIAA. This pattern highlights a critical concern for future PCS development strategies, particularly for EV owners in underprivileged communities. Individuals from these communities may encounter difficulties in finding conveniently located PCSs, potentially incurring an "invisible charging tax" through lost opportunities while waiting for their vehicles to charge. In contrast, EV users from other communities might benefit from the convenience of performing other activities while their vehicle charges, thereby making more efficient use of their time.

\subsection{Income-level Association with Accessibility Metrics}
Our previous findings underscore the necessity of investigating CBGs' income levels' association with gravity-based accessibility and the introduced opportunity-centric measures. This is partly due to the observed disparities in the opportunity-centric accessibility to the charging infrastructure, particularly the broader inequality in AICA compared to CIAA and gravity access, which points to a compounded lack of access to PCSs and an unequal distribution of opportunities in the metro areas. Moreover, the pronounced disparity in access to elastic activities as the primary travel choice and the inelastic activities as the secondary travel choice when selecting a charging station suggests potential structural barriers that may disproportionately affect certain populations, presenting challenges under ``invisible charging barrier" due to inconvenient charging options. 

\begin{figure}
	\centering\includegraphics[width=\textwidth]{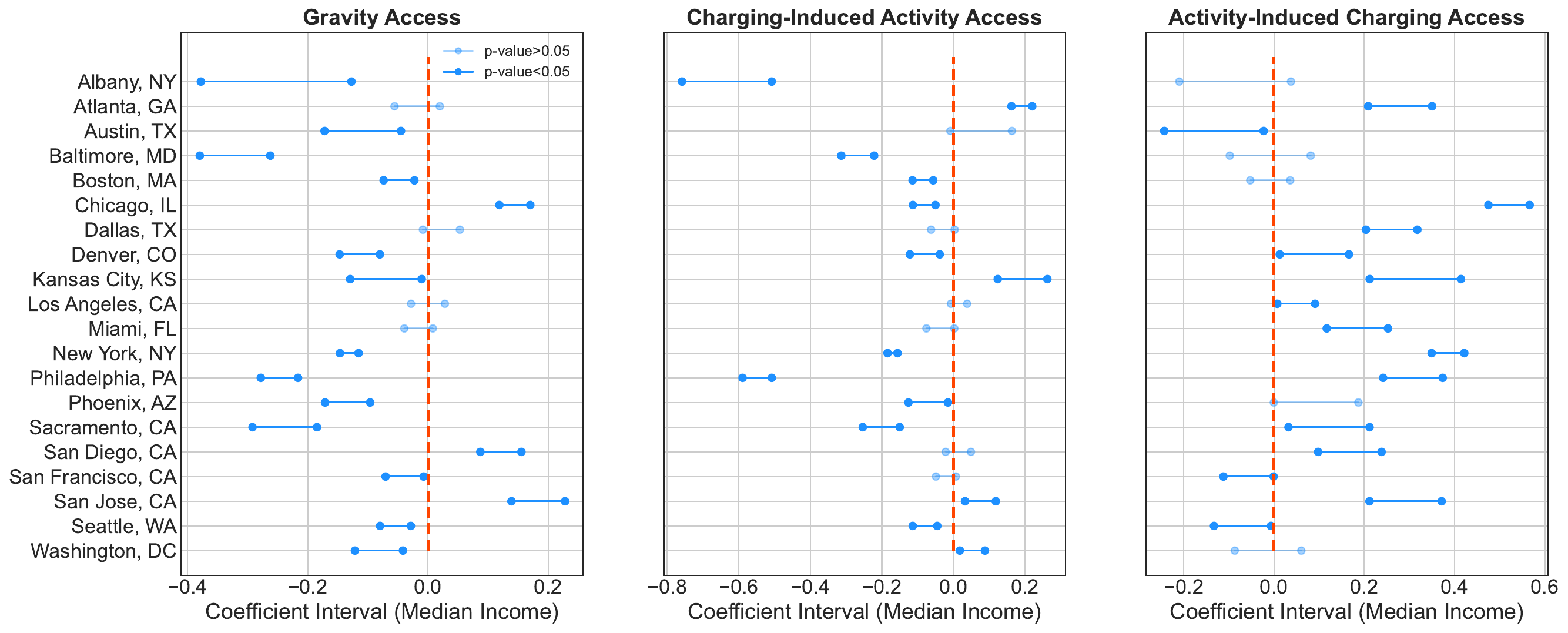}
	\caption{CBGs' Income Association with Accessibility Metrics}
	\label{fig:ols_income_metrics}
\end{figure}

Therefore, we assess the relationship between the median income levels of CBGs within metro areas (X) and their accessibility scores to charging infrastructure (Y) by performing an Ordinary Least Squares (OLS) regression, expressed as $ln(\gamma)\sim \beta ln(X)$, with R-squared values up to 0.17 (Albany, NY). The results, depicted in~\Cref{fig:ols_income_metrics}, showcase the standardized coefficient ($\beta$) intervals for median income level (X) at 0.05 significance level across different accessibility metrics. We report distinct disparities across accessibility metrics regarding the median income coefficients. In 13 of the 20 metropolitan areas, a negative and statistically significant relationship (p-value < 0.05) is observed between income levels and gravity-based accessibility to charging stations. For three metros-Chicago, San Diego, and San Jose -income was positively associated with gravity accessibility at 0.05 level. This implies that higher accessibility is often associated with lower income levels when distance is the only criterion for measuring access. We hypothesize this is due to the location of lower-income CBGs near downtown areas and city centers and the pattern of higher-income neighborhoods located in the suburbs. Moreover, when the focus shifts to Charging-induced Activity Accessibility (CIAA), where charging station access is the primary travel decision, the number of metros with a negative income coefficient decreases to 10, and metros with a positive significant relationship between income and accessibility increase to four (that is Atlanta, Kansas City, San Jose, and Washington D.C.). Since the CIAA already captures the distance impedance to the charging stations and only then adds a layer of access to nearby opportunities, we expect this modest change in coefficients to be a result of a high correlation between the gravity access and CIAA, ranging from 0.22 for Washington D.C. to 0.69 for Phoenix. In contrast, when examining Activity-induced Charging Accessibility (AICA), where the availability of activities takes a selection priority over the location of charging stations, a reverse trend is noted. Coefficients of income on AICA for 12 metros are significant and positive at the 0.05 level, with only 3 metros (Austin, San Francisco, and Seattle) showing a negative and statistically significant association at this level. 
These findings underscore an underlying structural inequality within metro areas, which can be at least partially linked to income distribution. Additionally, we highlight that gravity-based metrics alone fail to fully capture the nuances of charging station accessibility and disparities associated with it, emphasizing the importance of considering a broader range of opportunities nearby and the charging station selection process.      

\FloatBarrier
\FloatBarrier

\subsection{Counterfactual Analyses}
\begin{figure}
    \centering
    \begin{subfigure}[b]{\textwidth}
        \centering
        \includegraphics[width=\textwidth]{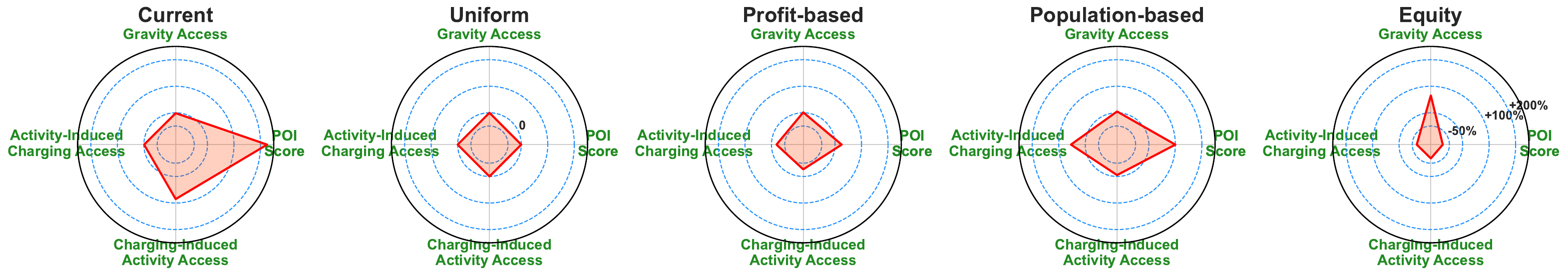}
        \caption{Scenarios Performance Compared to Uniform Deployment}
        \label{fig:radar_metrics}
    \end{subfigure}    
    \begin{subfigure}[b]{\textwidth}
        \centering
        \includegraphics[width=\textwidth]{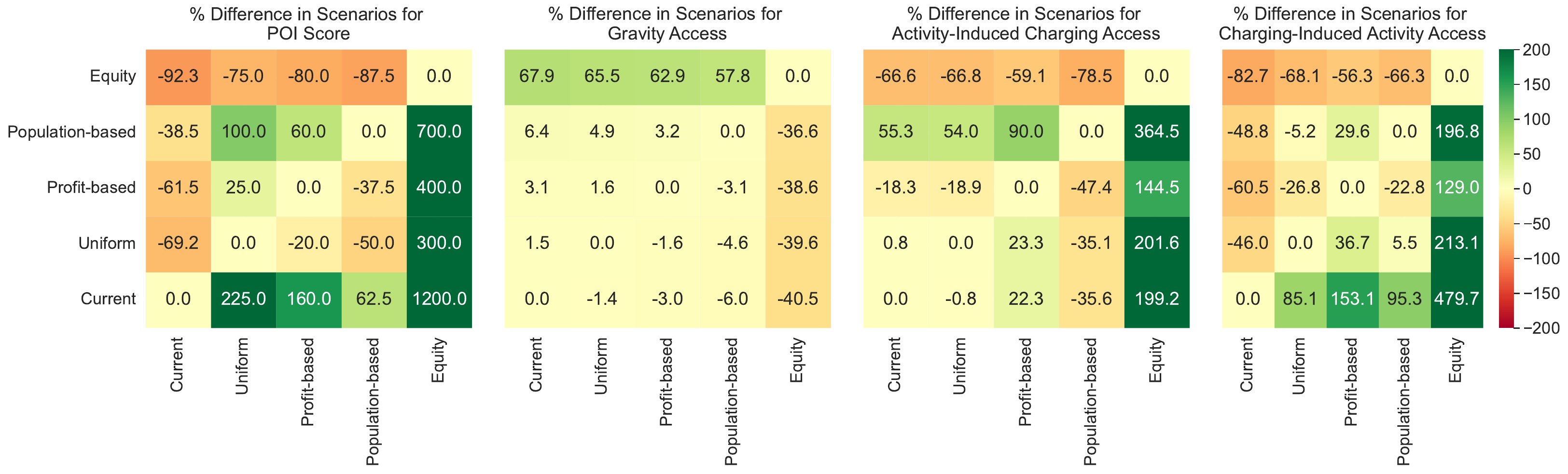}
        \caption{Scenarios Performance Comparison for each Accessibility Metric}
        \label{fig:heat_metrics}
    \end{subfigure}
    \caption{Performance Comparison of Different Deployment Scenarios}
    \label{fig:radar_heat}
\end{figure}

With the above discussions, we present our final results on the outcomes from the counterfactual analyses. \Cref{fig:radar_heat} illustrates the results of comparing the existing layout with alternative deployment strategies. We use radar plots to show the median performances per each metric across all metro areas, and the comparisons between the current deployment layout and four other alternative strategies are shown in~\Cref{fig:radar_metrics}. The results in the figure are scaled performances compared to the uniform strategy, with a value of 0 being the uniform baseline and the innermost ring representing 50\% decrease, with each outer ring representing an additional 100\% improvement over the uniform baseline. As an example, equity-based PCS deployment results in the worst CIAA performance (68\% lower than baseline), whereas the current PCS deployment scores 85\% greater median CIAA. Unsurprisingly, the equity-based approach obtains the highest PCS accessibility (65\% higher than the baseline), and current layout distribution leads to the worst performances in this category, but only with a 1.4\% decrease from the baseline strategy. At the same time, the current layout hold the highest POI scores and CIAA measures with levels of AICA accessibility performances near the baseline strategy. The POI scores for the current deployment are nearly 225\% better than the baseline (Uniform). Nevertheless, the AICA measure for the current deployment shows no improvement compared to the baseline (Uniform), where they are both 35\% lower than the population-based AICA median scores. It is worth stressing that the Equity-based approach, despite yielding the best accessibility measures, results in the worst performances for the opportunity scores, AICA, and CIAA measures, with over 67\% decrease compared to the baseline in all these metrics. This is key results as our addressing of equitable charging infrastructure deployment still relies on a broad set of tools centered around the geospatial location of underserved communities~\cite{Argonne2022, lee2022evi, zhou2022using}. We note that in this case, promoting a PCS deployment strategy that solely relies on equal distance to stations does not lead to equitable outcomes if the need to participate in additional activities while charging is not considered. We further observe while the current deployment suffers apparent disparities in each city, it still maintains a balance among all four measures. We conjecture that the disparities observed for current deployment may be due to the structural disparities that carry over from existing civil infrastructures (e.g., the road network and community segregation), which will require further investigation. 

\FloatBarrier
\FloatBarrier

\begin{figure}
    \centering
    \begin{subfigure}[b]{\textwidth}
        \centering
        \includegraphics[width=\textwidth]{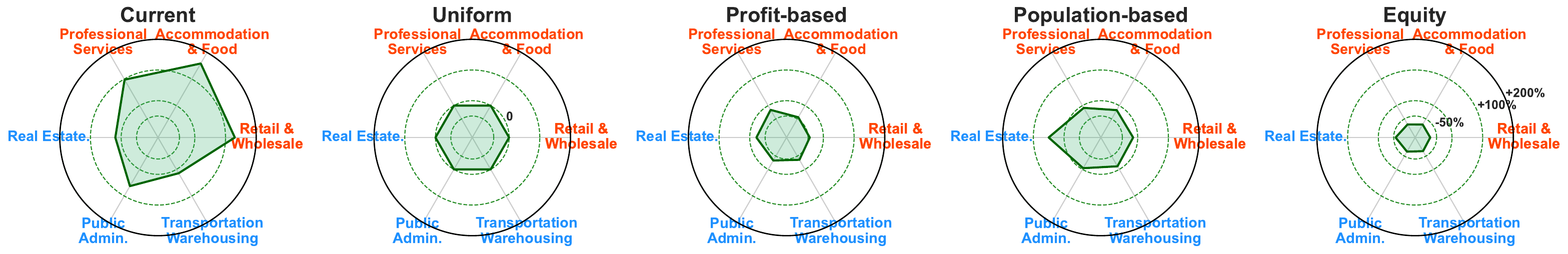}
        \caption{Charging Induced Activity Opportunities}
        \label{fig:radar_cia}
    \end{subfigure}    
    \begin{subfigure}[b]{\textwidth}
        \centering
        \includegraphics[width=\textwidth]{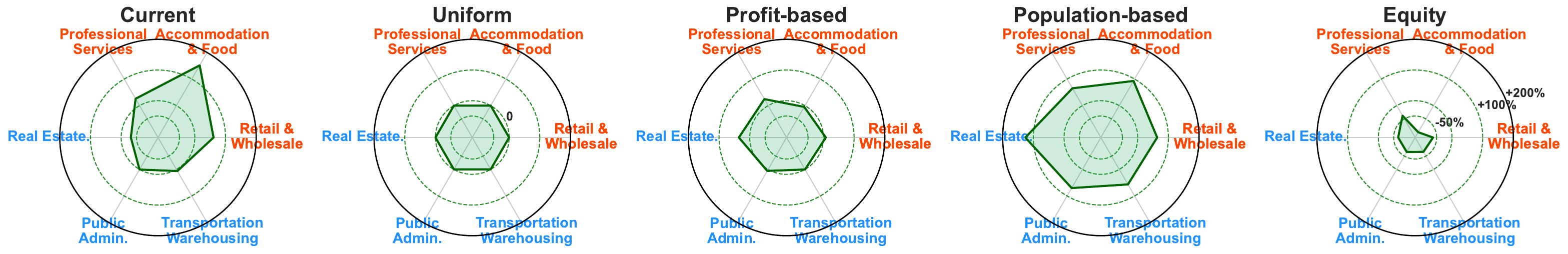}
        \caption{Activity Induced Charging Opportunities}
        \label{fig:radar_aic}
    \end{subfigure}
    \caption{Opportunity Measures Score of Different Deployment Scenarios}
    \label{fig:radar_aic_cia}
\end{figure}

While the above results provide an aggregate view of the CIAA and AICA performances, we further break them down into six individual POI categories as shown in~\Cref{fig:radar_cia} for the category-specific CIAA performances and in~\Cref{fig:radar_aic} for the category-specific AICA performances. Similar to the overall observations, we again confirm that the equity-based approach leads to the worst AICA and CIAA performances in every category of POIs with a reduction of at least 38\% compared to the baseline strategy. Based on the results, one important observation is that the current deployment strategy scores well in CIAA and AICA measures for elastic activity categories but yields worse performances for inelastic activity measures. We believe this is mainly due to the current policy guidance that promotes public-private partnerships in deploying PCSs and many local businesses investing in PCSs at their locations, hence the significantly higher CIAA measures, ranging from 14\% increase in `Transportation and Warehousing' to 159\% in `Accommodation and Food Services'. Nevertheless, the outcomes do not necessarily align with real-world needs if we consider the charging events being derived mainly from daily activities where the population-based approach gives the best overall AICA performance across the categories. This further highlights the disparities that arise from existing opportunity distributions in major metro areas. As discussed earlier, the nearby activity locations are more adjacent to charging opportunities for the more economically advantaged communities in terms of physical distance and the number of opportunities. This, together with the disparity in existing civil infrastructure, renders the planning of PCSs a highly challenging problem. Thus, we raise a concern that optimizing PCS deployment purely based on distance-based accessibility or opportunity-based accessibility will lead to undesirable outcomes that may either result in a waste of resources or exacerbate existing barriers in the community. It is important to balance the consideration across multiple dimensions and plan PCS proactively with the consideration of existing disparities and societal barriers, as discussed in this study, such that a PCS deployment plan for long-term community benefits can be achieved. 

\FloatBarrier
\FloatBarrier

Finally, we summarize the performance of the current PCS infrastructure in our counterfactual analysis, where we observe an increase in the median score value by at least 63\% for POI attachment and CIAA compared to all alternative deployment strategies. This further underscores the significance of activities nearby in locating the PCSs. However, as discussed earlier, a significant disparity in access to nearby activities remains, with over 50\% of cumulative POI scores concentrated around fewer than 10\% of the stations. Moreover, despite a low variance in gravity access scores across different strategies—excluding the equity-based approach—the current deployment yields the lowest scores with a median score 6\% lower than the population-based scenario. This highlights an underlying preference for locating stations, which results in reduced gravity accessibility (higher distance impedance) even compared to a uniform plan. Finally, the current strategy's AICA score is marginally lower than that of a uniform deployment (0.8\% lower) and 35\% less than the population-based deployment plan. This discrepancy is mainly due to the distribution skew within metro areas, where a few communities have high access to charging options when engaging in nearby activities while others face lower AICA scores. We previously showed this relationship to be significantly associated with income for 60\% of the metro areas. Overall, while the existing charging layout outperforms the alternative scenarios in two of the metrics, it also points to a need for a more equitable PCS deployment approach that ensures both the access to charging facilities and associated activities are more evenly distributed across communities.
\section{Discussion \& Conclusion}\label{section:discussion}
In this study, we incorporate an opportunity-centric aspect to the traditional view of civil infrastructure deployment, which enables us to connect the links between communities, existing charging layouts, and the location of activities. Our study allows for a thorough comparison of the existing charging infrastructure with various alternative charging deployment philosophies that might each prioritize a different aspect in addressing the complexity of the multi-faceted charging deployment problem. Our comprehensive analyses lead to four significant conclusions, outlined as follows:
\begin{enumerate}
    \item \textbf{There are notable disparities in how accessible charging stations are for different communities}. We observe that the tails of the CCDF for accessibility distributions across communities can be well approximated by the power-law distribution. This is indicative of a few communities having significantly better access to PCS than the majority of the communities. For instance, in San Jose, the PCS accessibility in the top 1\% of CBGs can be 3.6 times better than in the bottom 50\%. Meanwhile, we observe a highly skewed PCS score in the current layout, with less than 10\% of the PCSs accounting for more than 50\% of the cumulative share of POIs nearby across all 20 metro areas. This raises a critical concern where few PCSs are adjacent to the majority of the accessible activities, potentially exacerbating the inequalities in access when considering the activities.
    
    \item \textbf{Opportunity-based accessibility reveals more severe inequality issues for current PCSs}. 
    When we consider the intertwined relationship between activities and choices for charging, we notice that disparities in opportunity-based access within and between metropolitan areas are made worse. This underscores the inherent biases in where we position charging stations relative to activity hubs, which typically lean toward favoring specific areas. As a result, we see a more intense inequality in accessibility when activity locations are the primary decision factor, and charging stations are secondary. This outcome is particularly highlighted where, in the majority of the metro areas, CBGs with higher income are associated with better charging accessibility, 60\% being significant. This motivates us to reassess the quality of the existing charging infrastructure plan, as well as compare scenarios that prioritize different planning philosophies.   
    \item \textbf{Equitable deployment without considering the activities is not necessarily equitable}. Based on the results from counterfactual analyses, we report that a PCS deployment strategy that prioritizes an equitable conventional accessibility measure will result in the least equitable deployment outcomes with the worst opportunity-based accessibility metrics (at least 67\% decrease from a uniform deployment scenario). This finding is contrary to commonly recommended practices in other civil infrastructure deployments when distance-based accessibility is often used to evaluate the quality of the infrastructure deployment. This outcome highlights the complexity of planning PCSs as compared to other public infrastructures and urges the adoption of multidimensional metrics to guide the deployment of PCSs.
    
    \item \textbf{PCS deployment extends the existing inequalities in our cities.}
    The current PCS deployment strategy achieves a POI attachment and CIAA score improvement across all 20 metros over alternative strategies, with a median score increase of at least 63\%. Nevertheless, the existing layout results in the lowest gravity access scores, 6\% below the population-based scenario, indicating a preference for station location that compromises broader accessibility. Furthermore, the AICA score under the current deployment is slightly lower than uniform deployment by 0.8\% and significantly lower by 35\% compared to a population-based plan, reflecting a skewed distribution that favors a minority of higher-income communities for access to charging options while engaging in nearby activities, across 60\% of the metro areas. Overall, proactive planning of PCS and activity locations is essential to overturn this process, emphasizing distributing new PCSs more evenly rather than excessively attaching new PCSs to already-popular locations.
\end{enumerate}

Lastly, we suggest two future research directions that could enhance our study's findings. 
First, our study is able to link the disparities in opportunity-centric accessibility with variations in income levels across metropolitan areas. However, we hypothesize income is not the sole factor, and other potential barriers, including race and education, should also be taken into account to fully understand the complexities of accessibility issues and reveal if consistent patterns exist across the major metro areas.
Second, while our study employs accessibility as a static metric to explore the spatial relationship between communities and public infrastructure, incorporating data on population movements could offer more detailed insights. This could help us better understand location-specific measures of PCSs and potentially link to EV ownership across various communities. The results of such research could then be employed to jointly analyze and craft policies that promote long-term sustainability, considering both EV users and PCSs concurrently.

\printcredits

\section*{Declaration of competing interest}
The authors declare that they have no known competing financial interests or personal relationships that could have appeared to influence the work reported in this paper.

\section*{Acknowledgement}
The authors are grateful to the National Science Foundation for the award BCS \#2323732 to support the research presented in the paper. However, the authors are solely responsible for the findings presented in this study.
\bibliographystyle{abbrv}
\bibliography{cas-refs}

\begin{thebibliography}{10}

\bibitem{safegraphSpend}
{Advan}.
\newblock {A}dvan {M}onthly {P}atterns.
\newblock \url{https://advanresearch.com/}, 2023.

\bibitem{andrenacci2021modelling}
N.~Andrenacci, F.~Karagulian, and A.~Genovese.
\newblock Modelling charge profiles of electric vehicles based on charges data.
\newblock {\em Open Research Europe}, 1, 2021.

\bibitem{ardeshiri2020willingness}
A.~Ardeshiri and T.~H. Rashidi.
\newblock Willingness to pay for fast charging station for electric vehicles with limited market penetration making.
\newblock {\em Energy Policy}, 147:111822, 2020.

\bibitem{Argonne2022}
{Argonne National Laboratory}.
\newblock {Electric Vehicle Charging Equity Considerations}.
\newblock \url{https://www.anl.gov/esia/electric-vehicle-charging-equity-considerations}, 2022.

\bibitem{arlt2023retail}
M.-L. Arlt and N.~Astier.
\newblock Do retail businesses have efficient incentives to invest in public charging stations for electric vehicles?
\newblock {\em Energy Economics}, page 106777, 2023.

\bibitem{babar2023recharging}
Y.~Babar and G.~Burtch.
\newblock Recharging retail: Estimating consumer demand spillovers from electric vehicle charging stations.
\newblock {\em Boston University Questrom School of Business Research Paper}, 2023.

\bibitem{bao2021optimal}
Z.~Bao and C.~Xie.
\newblock Optimal station locations for en-route charging of electric vehicles in congested intercity networks: A new problem formulation and exact and approximate partitioning algorithms.
\newblock {\em Transportation Research Part C: Emerging Technologies}, 133:103447, 2021.

\bibitem{blinkcharging}
{Blink Charging}.
\newblock {E}lectric {V}ehicle {C}harging {T}ime {C}alculator.
\newblock \url{https://blinkcharging.com/drivers/electric-vehicle-time-charging-calculator}, 2024.

\bibitem{bruckmann2023experimental}
G.~Br{\"u}ckmann and T.~Bernauer.
\newblock An experimental analysis of consumer preferences towards public charging infrastructure.
\newblock {\em Transportation Research Part D: Transport and Environment}, 116:103626, 2023.

\bibitem{carley2019evolution}
S.~Carley, S.~Siddiki, and S.~Nicholson-Crotty.
\newblock Evolution of plug-in electric vehicle demand: Assessing consumer perceptions and intent to purchase over time.
\newblock {\em Transportation Research Part D: Transport and Environment}, 70:94--111, 2019.

\bibitem{carlton2022electric}
G.~J. Carlton and S.~Sultana.
\newblock Electric vehicle charging station accessibility and land use clustering: A case study of the chicago region.
\newblock {\em Journal of Urban Mobility}, 2:100019, 2022.

\bibitem{chen2020optimal}
R.~Chen, X.~Qian, L.~Miao, and S.~V. Ukkusuri.
\newblock Optimal charging facility location and capacity for electric vehicles considering route choice and charging time equilibrium.
\newblock {\em Computers \& Operations Research}, 113:104776, 2020.

\bibitem{clauset2009power}
A.~Clauset, C.~R. Shalizi, and M.~E. Newman.
\newblock Power-law distributions in empirical data.
\newblock {\em SIAM review}, 51(4):661--703, 2009.

\bibitem{do2018equitable}
B.~Do~Chung, S.~Park, and C.~Kwon.
\newblock Equitable distribution of recharging stations for electric vehicles.
\newblock {\em Socio-Economic Planning Sciences}, 63:1--11, 2018.

\bibitem{dong2006moving}
X.~Dong, M.~E. Ben-Akiva, J.~L. Bowman, and J.~L. Walker.
\newblock Moving from trip-based to activity-based measures of accessibility.
\newblock {\em Transportation Research Part A: policy and practice}, 40(2):163--180, 2006.

\bibitem{gastwirth1972estimation}
J.~L. Gastwirth.
\newblock The estimation of the lorenz curve and gini index.
\newblock {\em The review of economics and statistics}, pages 306--316, 1972.

\bibitem{ge2021there}
Y.~Ge, C.~Simeone, A.~Duvall, and E.~Wood.
\newblock There's no place like home: residential parking, electrical access, and implications for the future of electric vehicle charging infrastructure.
\newblock Technical report, National Renewable Energy Lab.(NREL), Golden, CO (United States), 2021.

\bibitem{globisch2019consumer}
J.~Globisch, P.~Pl{\"o}tz, E.~D{\"u}tschke, and M.~Wietschel.
\newblock Consumer preferences for public charging infrastructure for electric vehicles.
\newblock {\em Transport Policy}, 81:54--63, 2019.

\bibitem{guida2020urban}
C.~Guida and M.~Caglioni.
\newblock Urban accessibility: the paradox, the paradigms and the measures. a scientific review.
\newblock {\em TeMA-Journal of Land Use, Mobility and Environment}, 13(2):149--168, 2020.

\bibitem{guo2021disparities}
S.~Guo and E.~Kontou.
\newblock Disparities and equity issues in electric vehicles rebate allocation.
\newblock {\em Energy Policy}, 154:112291, 2021.

\bibitem{guo2020impacts}
Y.~Guo and S.~Peeta.
\newblock Impacts of personalized accessibility information on residential location choice and travel behavior.
\newblock {\em Travel Behaviour and Society}, 19:99--111, 2020.

\bibitem{guo2022modeling}
Y.~Guo, X.~Qian, T.~Lei, S.~Guo, and L.~Gong.
\newblock Modeling the preference of electric shared mobility drivers in choosing charging stations.
\newblock {\em Transportation Research Part D: Transport and Environment}, 110:103399, 2022.

\bibitem{hardman2018review}
S.~Hardman, A.~Jenn, G.~Tal, J.~Axsen, G.~Beard, N.~Daina, E.~Figenbaum, N.~Jakobsson, P.~Jochem, N.~Kinnear, et~al.
\newblock A review of consumer preferences of and interactions with electric vehicle charging infrastructure.
\newblock {\em Transportation Research Part D: Transport and Environment}, 62:508--523, 2018.

\bibitem{hsu2021public}
C.-W. Hsu and K.~Fingerman.
\newblock Public electric vehicle charger access disparities across race and income in california.
\newblock {\em Transport Policy}, 100:59--67, 2021.

\bibitem{ieaGlobalOutlook}
IEA.
\newblock {G}lobal {E}{V} {O}utlook 2023 – {A}nalysis.
\newblock \url{https://www.iea.org/reports/global-ev-outlook-2023}, 2023.
\newblock License: CC BY 4.0.

\bibitem{iravani2022multicriteria}
H.~Iravani.
\newblock A multicriteria gis-based decision-making approach for locating electric vehicle charging stations.
\newblock {\em Transportation Engineering}, 9:100135, 2022.

\bibitem{jia2023investigating}
W.~Jia and T.~D. Chen.
\newblock Investigating heterogeneous preferences for plug-in electric vehicles: Policy implications from different choice models.
\newblock {\em Transportation Research Part A: Policy and Practice}, 173:103693, 2023.

\bibitem{khan2022inequitable}
H.~A.~U. Khan, S.~Price, C.~Avraam, and Y.~Dvorkin.
\newblock Inequitable access to ev charging infrastructure.
\newblock {\em The Electricity Journal}, 35(3):107096, 2022.

\bibitem{knezovic2016enhancing}
K.~Knezovi{\'c}, S.~Martinenas, P.~B. Andersen, A.~Zecchino, and M.~Marinelli.
\newblock Enhancing the role of electric vehicles in the power grid: Field validation of multiple ancillary services.
\newblock {\em IEEE Transactions on Transportation Electrification}, 3(1):201--209, 2016.

\bibitem{lee2022evi}
D.-Y.~D. Lee, F.~Yang, A.~Wilson, and E.~Wood.
\newblock Evi-equity.
\newblock Technical report, National Renewable Energy Lab.(NREL), Golden, CO (United States), 2022.

\bibitem{lee2018charging}
H.~Lee and A.~Clark.
\newblock Charging the future: Challenges and opportunities for electric vehicle adoption.
\newblock 2018.

\bibitem{lei2022understanding}
T.~Lei, S.~Guo, X.~Qian, and L.~Gong.
\newblock Understanding charging dynamics of fully-electrified taxi services using large-scale trajectory data.
\newblock {\em Transportation Research Part C: Emerging Technologies}, 143:103822, 2022.

\bibitem{levinson2020towards}
D.~Levinson and H.~Wu.
\newblock Towards a general theory of access.
\newblock {\em Journal of Transport and Land Use}, 13(1):129--158, 2020.

\bibitem{li2022spatial}
G.~Li, T.~Luo, and Y.~Song.
\newblock Spatial equity analysis of urban public services for electric vehicle charging—implications of chinese cities.
\newblock {\em Sustainable Cities and Society}, 76:103519, 2022.

\bibitem{li2022subway}
X.~Li, G.~Xing, X.~Qian, Y.~Guo, W.~Wang, and C.~Cheng.
\newblock Subway station accessibility and its impacts on the spatial and temporal variations of its outbound ridership.
\newblock {\em Journal of Transportation Engineering, Part A: Systems}, 148(12):04022106, 2022.

\bibitem{liang2023effects}
J.~Liang, Y.~Qiu, P.~Liu, P.~He, and D.~L. Mauzerall.
\newblock Effects of expanding electric vehicle charging stations in california on the housing market.
\newblock {\em Nature Sustainability}, pages 1--10, 2023.

\bibitem{liao2017consumer}
F.~Liao, E.~Molin, and B.~van Wee.
\newblock Consumer preferences for electric vehicles: a literature review.
\newblock {\em Transport Reviews}, 37(3):252--275, 2017.

\bibitem{loni2023data}
A.~Loni and S.~Asadi.
\newblock Data-driven equitable placement for electric vehicle charging stations: Case study san francisco.
\newblock {\em Energy}, 282:128796, 2023.

\bibitem{luo2012variable}
W.~Luo and T.~Whippo.
\newblock Variable catchment sizes for the two-step floating catchment area (2sfca) method.
\newblock {\em Health \& place}, 18(4):789--795, 2012.

\bibitem{mcgrail2009measuring}
M.~R. McGrail and J.~S. Humphreys.
\newblock Measuring spatial accessibility to primary care in rural areas: Improving the effectiveness of the two-step floating catchment area method.
\newblock {\em Applied Geography}, 29(4):533--541, 2009.

\bibitem{nazari2022toward}
M.~Nazari-Heris, A.~Loni, S.~Asadi, and B.~Mohammadi-ivatloo.
\newblock Toward social equity access and mobile charging stations for electric vehicles: A case study in los angeles.
\newblock {\em Applied Energy}, 311:118704, 2022.

\bibitem{palmer2018total}
K.~Palmer, J.~E. Tate, Z.~Wadud, and J.~Nellthorp.
\newblock Total cost of ownership and market share for hybrid and electric vehicles in the uk, us and japan.
\newblock {\em Applied energy}, 209:108--119, 2018.

\bibitem{pan2019potential}
S.~Pan, A.~Roy, Y.~Choi, E.~Eslami, S.~Thomas, X.~Jiang, and H.~O. Gao.
\newblock Potential impacts of electric vehicles on air quality and health endpoints in the greater houston area in 2040.
\newblock {\em Atmospheric Environment}, 207:38--51, 2019.

\bibitem{pierce2023home}
L.~Pierce and P.~Slowik.
\newblock Home charging access and the implications for charging infrastructure costs in the united states.
\newblock 2023.

\bibitem{qian2019stationary}
X.~Qian, J.~Xue, S.~Sobolevsky, C.~Yang, and S.~Ukkusuri.
\newblock Stationary spatial charging demand distribution for commercial electric vehicles in urban area.
\newblock In {\em 2019 IEEE intelligent transportation systems conference (ITSC)}, pages 220--225. IEEE, 2019.

\bibitem{ratcliffe2016defining}
M.~Ratcliffe, C.~Burd, K.~Holder, and A.~Fields.
\newblock Defining rural at the us census bureau.
\newblock {\em American community survey and geography brief}, 1(8):1--8, 2016.

\bibitem{roy2022examining}
A.~Roy and M.~Law.
\newblock Examining spatial disparities in electric vehicle charging station placements using machine learning.
\newblock {\em Sustainable cities and society}, 83:103978, 2022.

\bibitem{saadati2022effect}
R.~Saadati, J.~Saebi, and M.~Jafari-Nokandi.
\newblock Effect of uncertainties on siting and sizing of charging stations and renewable energy resources: A modified capacitated flow-refueling location model.
\newblock {\em Sustainable Energy, Grids and Networks}, 31:100759, 2022.

\bibitem{safegraph_pois}
{SafeGraph}.
\newblock Evaluating safegraph data.
\newblock \url{https://docs.safegraph.com/docs/places-data-evaluation#section-recall-poi-count-per-category}, 2023.

\bibitem{safegraphGlobalPoints}
{SafeGraph}.
\newblock {G}lobal {P}oints of {I}nterest ({P}{O}{I}) {D}ata.
\newblock \url{https://www.safegraph.com/products/places}, 2023.

\bibitem{dotHistoricStep}
{The Federal Highway Administration}.
\newblock {H}istoric {S}tep: {A}ll {F}ifty {S}tates {P}lus {D}.{C}. and {P}uerto {R}ico {G}reenlit to {M}ove {E}{V} {C}harging {N}etworks {F}orward, {C}overing 75,000 {M}iles of {H}ighway.
\newblock \url{https://highways.dot.gov/newsroom/historic-step-all-fifty-states-plus-dc-and-puerto-rico-greenlit-move-\\ev-charging-networks}, 2022.

\bibitem{whitehouseFACTSHEET}
{The White House}.
\newblock {F}{A}{C}{T} {S}{H}{E}{E}{T}: {B}iden-{H}arris {A}dministration {A}nnounces {N}ew {S}tandards and {M}ajor {P}rogress for a {M}ade-in-{A}merica {N}ational {N}etwork of {E}lectric {V}ehicle {C}hargers.
\newblock \url{https://www.whitehouse.gov/briefing-room/statements-releases/2023/02/15/fact-sheet-biden-harris-administration-announces-new-standards-and-\\major-progress-for-a-made-in-america-national-network-of-electric-\\vehicle-chargers}, 2023.

\bibitem{toth2021inequality}
G.~T{\'o}th, J.~Wachs, R.~Di~Clemente, {\'A}.~Jakobi, B.~S{\'a}gv{\'a}ri, J.~Kert{\'e}sz, and B.~Lengyel.
\newblock Inequality is rising where social network segregation interacts with urban topology.
\newblock {\em Nature communications}, 12(1):1143, 2021.

\bibitem{naics2022}
{US Census Bureau}.
\newblock {N}orth {A}merican {I}ndustry {C}lassification {S}ystem ({NAICS}).
\newblock \url{https://www.census.gov/naics}, 2022.

\bibitem{censusAmericanCommunity}
{US Census Bureau}.
\newblock {A}merican {C}ommunity {S}urvey ({A}{C}{S}).
\newblock \url{https://www.census.gov/programs-surveys/acs}, 2023.

\bibitem{censusTIGERLineShapefiles}
{US Census Bureau}.
\newblock {T}{I}{G}{E}{R}/{L}ine {S}hapefiles.
\newblock \url{https://www.census.gov/geographies/mapping-files/time-series/geo/tiger-line-file.html}, 2023.

\bibitem{energyEEREAlternative}
{US Department of Energy}.
\newblock {E}{E}{R}{E}: {A}lternative {F}uels {D}ata {C}enter.
\newblock \url{https://afdc.energy.gov}, 2023.

\bibitem{uslu2021location}
T.~Uslu and O.~Kaya.
\newblock Location and capacity decisions for electric bus charging stations considering waiting times.
\newblock {\em Transportation Research Part D: Transport and Environment}, 90:102645, 2021.

\bibitem{wells2012converging}
P.~Wells.
\newblock Converging transport policy, industrial policy and environmental policy: the implications for localities and social equity.
\newblock {\em Local Economy}, 27(7):749--763, 2012.

\bibitem{xing2020environmental}
L.~Xing, Y.~Liu, B.~Wang, Y.~Wang, and H.~Liu.
\newblock An environmental justice study on spatial access to parks for youth by using an improved 2sfca method in wuhan, china.
\newblock {\em Cities}, 96:102405, 2020.

\bibitem{xu2022optimal}
D.~Xu, W.~Pei, and Q.~Zhang.
\newblock Optimal planning of electric vehicle charging stations considering user satisfaction and charging convenience.
\newblock {\em Energies}, 15(14):5027, 2022.

\bibitem{yang2023understanding}
Z.~Yang, X.~Li, Y.~Guo, and X.~Qian.
\newblock Understanding active transportation accessibility's impacts on polycentric and monocentric cities' housing price.
\newblock {\em Research in Transportation Economics}, 98:101282, 2023.

\bibitem{zhang2023electric}
R.~Zhang, N.~Horesh, E.~Kontou, and Y.~Zhou.
\newblock Electric vehicle community charging hubs in multi-unit dwellings: Scheduling and techno-economic assessment.
\newblock {\em Transportation Research Part D: Transport and Environment}, 120:103776, 2023.

\bibitem{zhou2022using}
Y.~Zhou, D.~Gohlke, M.~Sansone, J.~Kuiper, and M.~P. Smith.
\newblock Using mapping tools to prioritize electric vehicle charger benefits to underserved communities.
\newblock Technical report, Argonne National Lab.(ANL), Argonne, IL (United States), 2022.

\end{thebibliography}

\end{document}